\documentclass[onecolumn,superscriptaddress,floatfix,preprintnumbers, nofootinbib,hyperref]{revtex4-2} 
\pdfoutput=1
\usepackage[colorlinks=true,breaklinks=true]{hyperref}
\usepackage[normalem]{ulem}
\usepackage[utf8]{inputenc}
\hypersetup{allcolors=[rgb]{0.0 0.0 0.6},linkcolor=[rgb]{0.75 0.05 0.05}}
\usepackage{amsmath,amssymb}
\usepackage{epsfig}  
\usepackage{graphicx}   
\usepackage{slashed}       
\usepackage{tikz}
\usepackage{tikz-feynman}
\usepackage{braket}
\usepackage{url}
\usepackage{color}
\usepackage{multirow}
\usepackage{comment}
\usepackage{amssymb}
\usepackage{orcidlink}
\usepackage[version=3]{mhchem}
\DeclareUnicodeCharacter{2212}{-}
\usepackage{amsmath,amssymb}
\usepackage{epsfig}  
\usepackage{graphicx}   
\usepackage{slashed}
\usepackage{color}
\usepackage{url}
\usepackage{array}
\newcolumntype{L}[1]{>{\raggedright\let\newline\\\arraybackslash\hspace{0pt}}m{#1}}
\newcolumntype{C}[1]{>{\centering\let\newline\\\arraybackslash\hspace{0pt}}m{#1}}
\newcolumntype{R}[1]{>{\raggedleft\let\newline\\\arraybackslash\hspace{0pt}}m{#1}}
\usepackage{color}
\usepackage{multirow}
\usepackage{comment}
\usepackage{amssymb}
\usepackage{amsmath}

\DeclareMathOperator{\MeV}{MeV}

\DeclareMathOperator{\diag}{diag}

\DeclareMathOperator{\g}{g}

\DeclareMathOperator{\kton}{kton}

\DeclareUnicodeCharacter{2212}{-}

\newcommand{\bq}{{\bf q}}
\newcommand{\bp}{{\bf p}}

\newcommand{\beq}{\begin{equation}}
\newcommand{\eeq}{\end{equation}}

\newcommand{\beqn}{\begin{eqnarray}} 
\newcommand{\eeqn}{\end{eqnarray}} 
\newcommand{\bsigma}{\mbox{\boldmath $\sigma$}} 
 
\newcommand{\half}{\frac{1}{2}} 

\newcommand{\br}{{\bf r}}
\newcommand{\sumint}{\sum \!\!\!\!\!\!\!\!\int } % soimma + integrale
\newcommand{\barv}{\overline{V}}
\newcommand{\threej}[6]{ \left( \begin{array}{ccc} 
                               #1 & #2 & #3 \\ 
                               #4 & #5 & #6  
                             \end{array} 
                        \right) }

\begin{document}

\noindent 
 \title{Cross section for supernova axion observation\\ in neutrino water {\v C}herenkov detectors}

\author{Pierluca~Carenza}\email{pierluca.carenza@fysik.su.se}
\affiliation{The Oskar Klein Centre, Department of Physics, Stockholm University, Stockholm 106 91, Sweden
}

\author{Giampaolo~Co'}\email{Giampaolo.Co@le.infn.it>}
\affiliation{Dipartimento di Matematica e Fisica ``Ennio De Giorgi'', Universit\`a del Salento, Via Arnesano, 73100 Lecce, Italy}
\affiliation{Istituto Nazionale di Fisica Nucleare - Sezione di Lecce,
Via Arnesano, 73100 Lecce, Italy}

\author{Maurizio~Giannotti}
\email{mgiannotti@barry.edu}
\affiliation{Department of Chemistry and Physics, Barry University, 11300 NE 2nd Ave., Miami Shores, FL 33161, USA}%

\author{Alessandro~Lella}
\email{alessandro.lella@ba.infn.it}
\affiliation{Dipartimento Interateneo di Fisica  ``Michelangelo Merlin,'' Via Amendola 173, 70126 Bari, Italy}
\affiliation{Istituto Nazionale di Fisica Nucleare - Sezione di Bari, Via Orabona 4, 70126 Bari, Italy}%

\author{Giuseppe~Lucente}
\email{giuseppe.lucente@ba.infn.it}
\affiliation{Dipartimento Interateneo di Fisica  ``Michelangelo Merlin,'' Via Amendola 173, 70126 Bari, Italy}
\affiliation{Istituto Nazionale di Fisica Nucleare - Sezione di Bari, Via Orabona 4, 70126 Bari, Italy}%

\author{Alessandro~Mirizzi}
\email{alessandro.mirizzi@ba.infn.it}
\affiliation{Dipartimento Interateneo di Fisica  ``Michelangelo Merlin,'' Via Amendola 173, 70126 Bari, Italy}
\affiliation{Istituto Nazionale di Fisica Nucleare - Sezione di Bari, Via Orabona 4, 70126 Bari, Italy}%

 \author{Thomas~Rauscher}
 \email{Thomas.Rauscher@unibas.ch}
 \affiliation{Department of Physics, University of Basel, Klingelbergstr. 82, CH-4056 Basel, Switzerland}
 \affiliation{Centre for Astrophysics Research, University of Hertfordshire, Hatfield AL10 9AB, United Kingdom}
 
\date{\today}

\begin{abstract}
Axions coupled to nucleons might be copiously emitted from core-collapse supernovae (SNe). 
If the axion-nucleon coupling is strong enough, axions would be emitted from the SN as a burst and, reaching Earth, may excite the oxygen nuclei in water {\v C}herenkov detectors (${}^{16}{\rm O} + a \to  {}^{16}{\rm O}^{*}$). This process will be followed by decay(s) of the excited state resulting in an emission of photons and thus providing a possibility for a direct detection of axions from a Galactic SN in large underground neutrino {\v C}herenkov detectors. Motivated by this possibility, we present an updated calculation of axion-oxygen cross section obtained by using self-consistent continuum Random Phase
Approximation. We calculate the branching ratio of the oxygen nucleus de-excitation into gamma-rays, neutrons, protons and $\alpha$-particles and also consider photon emission from secondary nuclei to compute a total $\gamma$ spectrum created when axions excite ${}^{16}{\rm O}$.  
These results are used to revisit the detectability of axions from SN 1987A in \mbox{Kamiokande-II}. 
\end{abstract}
\maketitle

\section{Introduction}
\label{sec:intro}

Core-collapse supernovae (SNe) are recognized as powerful laboratories for particle physics~\cite{Raffelt:1999tx}. Notably, the SN neutrino burst has been observed once in coincidence with SN 1987A explosion~\cite{Kamiokande-II:1987idp,Hirata:1988ad,IMB:1988suc,Bionta:1987qt}. 
Despite the sparseness of the neutrino data, this detection is considered as a milestone in astroparticle physics allowing one not only to probe fundamental neutrino properties, but also to constrain novel particle emission~\cite{Raffelt:1987yt,Raffelt:1990yu}. 
Indeed, if \emph{weakly-coupled} 
exotic particles were emitted from a SN core together with  neutrinos, then they would reduce the energy budget available for neutrinos, shortening the duration of the observed burst. 

From the absence of a sizable reduction of the SN 1987A neutrino burst duration, several scenarios have been constrained.
These include the emission of axions~\cite{Raffelt:1987yt,Keil:1996ju,Chang:2018rso,Carenza:2019pxu,Carenza:2020cis}, scalar bosons~\cite{Caputo:2021rux}, sterile neutrinos~\cite{Kolb:1996pa,Raffelt:2011nc,Mastrototaro:2019vug}, dark photons~\cite{Chang:2016ntp}, light $CP$-even scalars~\cite{Dev:2020eam}, dark flavored particles~\cite{Camalich:2020wac} and unparticles~\cite{Hannestad:2007ys} (see Refs.~\cite{Mirizzi:316773,Antel:2023hkf} for a list).
These constraints are based on an indirect signature on the observable neutrino burst. However, if novel particles are \emph{strongly-coupled} with matter, then one would expect them to be trapped in the SN core, analogously as neutrinos, and to be emitted from their last-scattering surface producing a \emph{burst}. 
In this scenario one may wonder if such a burst can lead to an observable signal in large underground neutrino detectors.

In this context, the case of \emph{direct detection} of axions from SN has been studied in relation to the SN 1987A event. 
If the axion-nucleon coupling  $g_{aN}$ is strong enough, axions emitted during a SN explosion may lead to a detectable signal in  large water {\v C}herenkov neutrino detectors, as  proposed in the seminal paper by Engel \emph{et al.}~\cite{Engel:1990zd}. The authors of Ref.~\cite{Engel:1990zd} proposed to look for axion-induced excitation of oxygen nuclei with the subsequent emission of a photon to relax the system

%%%%%%%%%%%%%%%%%%%%%%%%%%%%%%%%%%%%%%%%%%%%%
\begin{equation}
    a+ {}^{16}{\rm O}\rightarrow {}^{16}{\rm O}^* \rightarrow \begin{cases}\,{}^{16}{\rm O}+\gamma\\
    \,{}^{15}{\rm O}+\mathrm{n}+\gamma\\
    \,{}^{15}{\rm N}+\mathrm{p}+\gamma\quad.
    \end{cases}
\end{equation}
%%%%%%%%%%%%%%%%%%%%%%%%%%%%%%%%%%%%%%%%%%%%%

If an axion burst had occurred during the SN 1987A explosion, then for sufficiently high values of $g_{aN}$ it would have been detected at Kamiokande II (KII) water {\v C}herenkov detector~\cite{Kamiokande-II:1987idp,Hirata:1988ad}, as pointed out in Ref. ~\cite{Engel:1990zd}. 
The absence of such a signal  places stringent constraints on the value of $g_{aN}$.

Given the recent interest in axion emission from SNe we find it timely to revisit the calculation of the axion-oxygen cross section by using refined nuclear models. 
There are various problems in carrying out this calculation. 
A first one is related to the fact that the energy range of strongly-coupled SN axions is $\mathcal{O}(10~\MeV)$. 
Axions excite the nucleus in a region where single-particle and collective excitation modes overlap.  Therefore, a theory describing both kinds of excitation in a unique framework is required. A second problem emerges because these excitation energies allow for the nucleon emission, which has to be considered in the theory describing the excitation of the nuclear system. A third problem is connected to the specific characteristics of the axion-nucleon coupling, allowing only nuclear unnatural parity states to be excited. Experimentally, in the region  of interest, these states are poorly known since their cross section is overwhelmed by the contribution of the natural parity states. Therefore, there is no experimental guide to set up a phenomenological nucleon-nucleon
interaction. 

We tackle these problems by describing the nuclear excitation with the continuum Random Phase Approximation (CRPA) theory which considers the excitation of the continuum spectrum, i.e. the one-nucleon emission. Furthermore, our approach is self-consistent, in the sense  that we use the same effective nucleon-nucleon interaction to describe the nuclear ground state as well as the excited states. Since this interaction has a finite range, it does not require any further renormalization parameter to stabilize  the CRPA results~\cite{don11a}. 

In this article we estimate the axion-nucleus cross section and evaluate the uncertainty related to the nuclear model. The plan of our work is as follows. In Section~\ref{sec:axnuc} we revise the axion-nucleon Lagrangian which is the starting point  of our calculations.
In Section~\ref{sec:axnucleus}  we show how we obtain the axion-nucleus cross section. 
In Section~\ref{sec:rpa}  we present our self-consistent nuclear model where the
ground state is described by using the Hartree-Fock (HF) and the excited states by the
CRPA.  In Section~\ref{sec:results1} we discuss the results obtained by calculating the axion-$^{16}$O cross section. In Section~\ref{sec:gammad} we apply these results  to calculate the gamma-ray spectrum from oxygen de-excitation. 
Then, in  Section~\ref{sec:results2} we estimate the expected axion event rate in the Kamiokande-II experiment under the assumption of a supernova similar to the SN 1987A explosion. Finally, in Section~\ref{sec:summary} we summarize our results and conclude.

%%%%%%%%%%%%%%%%%%%%%%%%%%%%%%%%
\section{Axion-nucleon interaction}
\label{sec:axnuc}

The axion interaction with nucleons is described by the following Lagrangian~\cite{Engel:1990zd}
%
%\begin{align}
\beq
\mathcal{L}_{aN} (\br,t) =
\frac{g_{aN}}{2m_{N}}\bar{\psi}_N(\br,t)
\gamma^{\mu}\gamma^{5}(C_{0}+C_{1}\tau_{3})\psi_N(\br,t) \partial_{\mu}a(\br,t)
\,,
 \label{eq:Lagrangian}
%\end{align}
\eeq
where $g_{aN}$ is the axion-nucleon coupling constant, $m_N$ is the nucleon mass, $\psi_N = (p, n)^T$ is the nucleon doublet, $a$ is the axion field, $\tau^3={\rm diag}(1,-1)$ is the third Pauli matrix, and $C_{0}=(C_{p}+C_{n})/2$ and $C_{1}=(C_{p}-C_{n})/2$ are the model-dependent iso-scalar and iso-vector couplings connected with the axion-proton $C_{p}$ and axion-neutron $C_{n}$ model-dependent constants.
For definiteness, in the following we will fix $C_p=-0.47$ and $C_n\simeq0$, as in the Kim-Shifman-Vainshtein-Zakharov (KSVZ) axion model~\cite{Kim:1979if,Shifman:1979if}. 

From a theoretical standpoint, 
the axion-nucleon interactions are particularly interesting since they receive an unavoidable contribution from the defining coupling of axions to gluons, just like the photon coupling. 
Furthermore, contrarily to the axion-photon coupling, it is theoretically difficult to suppress these couplings in axion models without spoiling the solution of the strong-$CP$ problem of Quantum Chromodynamics (QCD)~\cite{DiLuzio:2017ogq,DiLuzio:2020wdo,DiLuzio:2021ysg}. 
Hence, the detection of a pseudo-scalar particle with a sizable coupling to nucleons would be a strong indication of the QCD axion.

The phenomenology associated with the interaction in Eq.~\eqref{eq:Lagrangian} has been extensively explored in recent years (see, e.g., Refs.~\cite{Moriyama:1995bz,Krcmar:1998xn,CAST:2009jdc,CUORE:2012ymr,DiLuzio:2021qct,Borexino:2012guz,Bhusal:2020bvx,Lucente:2022esm,Alonso:2021kyu,Waites:2022tov}).
One of the most efficient laboratories to study these interactions is a SN explosion (see, e.g., Refs.~\cite{Raffelt:1987yt,Keil:1996ju,Chang:2018rso,Carenza:2019pxu,Carenza:2020cis}). 
In particular, the detection of neutrinos associated with SN 1987A has provided a valuable argument to constrain the SN axion production. 
For the purpose of the current paper, it is particularly interesting the case of axions so \emph{strongly-coupled} with matter to be trapped in the SN core, analogously to neutrinos. 
In this case, corresponding to $g_{aN}\gtrsim10^{-7}$~\cite{Lella:2023bfb},
axions thermalize and are emitted from their last-scattering surface producing a \emph{burst}. 
Our aim is to explore the possibility of detecting this burst in large neutrino underground {\v C}herenkov detectors due to the axion absorption on oxygen nuclei. 

The description of this process is highly non-trivial, involving the interaction of an elementary particle, the axion, with a many-body system. 
We first study the interaction between the axion and a single nucleon, assumed to be a structureless particle (see Sec.~\ref{sec:axnucleus}), and, in a second step, we consider the nucleus as a many-nucleon system (see Sec.~\ref{sec:rpa}).

%%%%%%%%%%%%%%%%%%%%%%%%
\section{Axion-nucleus cross section}
\label{sec:axnucleus}

We are interested in describing a process where an axion is absorbed by a nucleus which reaches an excited state. 
From the Lagrangian density of Eq.~(\ref{eq:Lagrangian}) we obtain the following Hamiltonian operator describing the interaction between axions and nucleons
\beq
\mathcal {H}_{aN}(\br,t) =
-\frac{g_{aN}}{2 m_N} \bar{\psi}_N(\br,t)
\gamma^{k}\gamma^{5}(C_{0}+C_{1}\tau_{3})  \psi_N(\br,t)
\partial_{k}a(\br,t)
\equiv  -\frac{g_{aN}}{2 m_N} \mathcal{J}^k(\br,t) \partial_{k}a(\br,t)
\,,
\label{eq:Hamiltonian}
\eeq
where the index $k$ assumes values from 1 to 3. The transition amplitude
describing this process can be expressed as
\beq
\mathcal{M}_{if} = \int d^3 \br\, dt \braket{f| \mathcal {H}_{aN}(\br,t)| i}
=  -\frac{g_{aN}}{2 m_N} \int d^3 \br\, dt \braket{\Psi_f| \mathcal{J}^k(\br,t) |\Psi_i}
\braket{0|\partial_{k}a(\br,t)|\bp}
\;,
\label{eq:mif}
\eeq
where we have separated the contribution of the hadronic current $\mathcal{J}^k$
from that of the axion field $a(\br,t)$. 
The state $\ket{\bp}$ describes an axion of momentum $\bp$, and $\ket{0}$
is the physical vacuum. We express the axion field in a plane
wave basis
\beq
a(\br,t) = \int \frac{d^3q}{\sqrt{2 \epsilon_q} } 
\left( 
e^{-i \epsilon_q t + i \bq \cdot \br} \, b_q
+ e^{i \epsilon_q t - i \bq \cdot \br} \, b^\dagger_q
\right)
\;,
\label{eq:afield}
\eeq
where $b_q$ and $b^\dagger_q$ are the axion
creation and destruction operators. In Eq.~\eqref{eq:afield} we assume a unitary volume of dimensions $[{\rm energy}]^{-3}$ to provide the axion field with the correct dimensions. Since we neglect the axion rest mass, we have $\epsilon_q=|\bq|$ in natural units. By inserting the expression of Eq.~(\ref{eq:afield}) in the axion matrix element term of Eq.~(\ref{eq:mif}),
 we obtain
\beq
\braket{0|\partial_{k}a(\br,t)|\bp} = 
i p_k \frac{e^{-i \epsilon_p t + i \bp \cdot \br}} {\sqrt{2 \epsilon_p}}
\label{eq:amatel}
\;\;.
\eeq
We express the nuclear matrix element of Eq.~(\ref{eq:mif}) in Heisenberg representation
\beq
\braket{\Psi_f| \mathcal{J}^k(\br,t) |\Psi_i} = 
\braket{\Psi_f| e^{i H t} \mathcal{J}^k(\br,0) e^{-iHt} |\Psi_i} 
= e^{i (E_f - E_i) t } \braket{\Psi_f| \mathcal{J}^k(\br) |\Psi_i}
\;,
\label{eq:heis}
\eeq
where $H$ is the Hamiltonian describing the nuclear system and 
$\Psi_i$ and $\Psi_f$ its eigenstates.
In the last step we have used the expression $\mathcal{J}^k(\br,0) \equiv \mathcal{J}^k(\br)$.

By inserting Eqs.~(\ref{eq:amatel}) and~(\ref{eq:heis}) into Eq.~(\ref{eq:mif}) and 
carrying out the time integration with the usual procedure, as indicated for example in 
Ref.~\cite{bjo64}, we obtain

\beq
\mathcal{M}_{if}= - \frac{g_{aN}}{2 m_N} 
\frac {2 \pi \delta(E_f - E_i - \epsilon_p)} {\sqrt{2 \epsilon_p }}
\int d^3 \br \,\braket{\Psi_f| \mathcal{J}^k(\br) |\Psi_i} 
\partial_k e^{i \bp \cdot \br}\,.
\label{eq:mif2}
\eeq
We carry out a multipole expansion of the exponential term in the equation above and, 
by choosing the $z$ axis along the direction of $\bp$ we have
\beq
e^{i \bp \cdot \br} = \sum_{l=0}^\infty i^l 
\sqrt{4 \pi (2l+1)} j_l(pr) Y_{l,0}(\Omega) \,,
\eeq
where $p\equiv|\bp|$ and $r\equiv|\br|$, $j_l$ is the spherical Bessel function,
$Y_{l,0}$ the spherical harmonic and $\Omega$ indicates the angular polar 
coordinates relative to the position of $\br$. By using this expression we can 
rewrite the matrix element in Eq.~(\ref{eq:mif2}) as
\beq
\mathcal{M}_{if}= \sum_{l=0}^\infty \mathcal{M}_{if}^{\,l}= -\frac{g_{aN}}{2 m_N} 
\frac {2 \pi \delta(\omega - \epsilon_p)} {\sqrt{2 \epsilon_p }} \epsilon_p \sum_{l=0}^\infty i^{l-1} \sqrt{4 \pi (2l+1)} L_{l,0}\,,
\label{eq:mif3}
\eeq
where $\omega=E_f-E_i$ is the nuclear excitation energy and $M_{if}^{\,l}$ stands for the 
$l$-th term of 
the sum. In this expression, $L_{l,0}$ is defined as:
\begin{equation}
L_{l,0}=\frac{i}{p}
    \int  d^{3}\br\,\braket{\Psi_f| \mathcal{J}^k(\br) |\Psi_i} 
    \partial_{k}\left[ j_{l}(pr)Y_{l,0}(\Omega)\right]
    \,.
\label{eq:Ll0}
\end{equation}
Notice that we have multiplied by $\epsilon_p$
and divided by $p$ which, in our assumption of axion zero mass, are the same quantity.
The nuclear states $\Psi$ are identified by a total angular momentum $J$ and
a parity $\Pi$. We are interested in even-even nuclei, specifically the $^{16}$O, 
whose ground states are always characterized by $J=0$ and $\Pi=+1$.
We can rewrite Eq.~(\ref{eq:Ll0}) as 
\beq
L_{l,0}=\frac{i}{p}
\braket{J M ;\Pi|  \int  d^{3}\br\, \mathcal{J}^k(\br) 
    \partial_{k}\left[ j_{l}(pr)Y_{l,0}(\Omega)\right] 
    | 0 0; +1}
    \equiv \braket{J M ; \Pi | T_{l,0} | 0\,0; +1}\,,
\eeq
where $J$, $M$ and $\Pi$ are the angular momentum, its third component and the parity of the final nuclear state, respectively.
We apply the Wigner-Eckart theorem~\cite{edm57} and obtain
\beq
L_{l,0}= (-1)^{J-M} \threej{J}{l}{0}{-M}{0}{0} \braket{J^\Pi || T_l || 0^+}\,,
\label{eq:wigner}
\eeq
where we used the Wigner 3-j symbol, and we indicate
with $\braket{J^\Pi || T_l || 0^+}$  a reduced matrix element~\cite{edm57}  for the angular part, 
and an integration on $r$. For the properties of the 3-j symbol we have $l=J$ and $M=0$. 

Because of the orthogonality properties of the 3-j symbol, there are no interference terms between
amplitudes with different angular momentum excitations. Therefore, the total cross section
can be expressed as
\cite{Engel:1990zd}
\begin{equation}
        \sigma(\epsilon_p)=\int \frac{d\omega }{\cal T} \sum_{J} 
        \left( \frac{d\rho}{d\omega} \right)_{J}
         \left |M_{fi}^{\,J}\right|^2
 =\frac {4 \pi^2 g^2_{aN}}{m^2_N}  \sum_{J}\,\epsilon_p\,(2J+1) \left| \braket{J^\Pi || T_J || 0^+}\right|^2\delta(\epsilon_p-E_{J})\,,
\label{eq:xsect2}
\end{equation}
where ${\cal T}$ is the observation time and we have exploited the property of the delta function $\delta^2(E)={\cal T}/2\pi\,\delta(E)$ in the limit ${\cal T}\rightarrow\infty$. Here, the squared matrix element $|M_{fi}^{\,J}|^2$ is obtained by averaging the transition amplitude over nucleon initial states and summing over final states while $\left( {d\rho}/{d\omega} \right)_{J}=4\,\delta(\omega-E_J)$ is the nuclear states density for the excitation level $J^{\Pi}$ with energy $E_J$.

\section{The nuclear model}
\label{sec:rpa}
The transition operator $T_{J,M}$ is obtained from the Hamiltonian in Eq.~(\ref{eq:Hamiltonian})
describing the axion-nucleon interaction, implying that $T_{J,M}$ is 
a one-body operator. Thus, the total axion-absorption cross section by a nucleus is
mediated by the interaction with a single nucleon, which is part of a nuclear system composed of interacting nucleons. 

In Appendix \ref{app:nuccurrent} we present a non-relativistic reduction of the 
nucleonic current $\mathcal{J}^k$ in Eq.~(\ref{eq:Hamiltonian}), leading to
the expression
\beq
\mathcal{J}^k(\br) \longrightarrow \sum_{i=1}^A
[C_0 + C_1 \tau_3(i)] \sigma^k \delta^3(\br - \br_i)
\;\;,
\label{eq:nucleonic}
\eeq
where the sum runs over all the $A$ nucleons. 
We express the transition matrix element in Eq.~(\ref{eq:xsect2}) as
\begin{equation}
  \braket{J^\Pi || T_J || 0^+} =   \frac{i}{\epsilon_{p}} \int d^3 \br \,
    \braket{J^{\Pi} ||  \sum_{i=1}^A 
  (C_{0}+C_{1}\tau_{3}(i)){\bsigma}\cdot{\mathbf{\nabla}}(j_{J}(pr)Y_{J,0}(\Omega))  || 0^+}
    \;. 
\end{equation}
The evaluation of this matrix element requires models 
describing the nuclear ground and excited states
in terms of nucleonic degrees of freedom.
The starting point is the definition of a basis of single-particle
wave functions related to each individual nucleon. 
In our calculations, this single particle basis is generated by 
solving a set of spherical HF equations and considering
effective nucleon-nucleon 
interactions containing spin-orbit terms \cite{bau99}, and, eventually,
also tensor terms \cite{co98b}. 

The excitation induced by the operator in Eq.~(\ref{eq:nucleonic}) is described by promoting nucleons lying on single-particle states below the Fermi surface (hole states $h$)  
to single-particle states above it (particle states $p$). 
In the present approach, we use the RPA theory for the description of the excited states, and 
leading to the following expression of the transition matrix element
\beqn
\nonumber
\braket{J^\Pi || T_J || 0^+}^{\rm RPA}
&=& 
\frac{i}{\epsilon_{p}} \int d^3 r \,
\sum_{p,h} \sum_{k=1}^A
\Big[X^{J^\Pi}_{ph} \braket{p ||   
(C_{0}+C_{1}\tau_{3}(k)){\bsigma}\cdot{\mathbf{\nabla}}(j_{J}(pr)Y_{J,0}(\Omega))  || h} \\
&~& - Y^{J^\Pi}_{ph} \braket{h ||   
(C_{0}+C_{1}\tau_{3}(k)){\bsigma}\cdot{\mathbf{\nabla}}(j_{J}(pr)Y_{J,0}(\Omega))  || p} \Big]
\;\;.
\label{eq:RPAmatel}
\eeqn
In this equation the RPA amplitudes $X$ and $Y$ are real
numbers obtained by solving the secular RPA equations \cite{rin80,co23}, 
and $\ket{p}$ and $\ket{h}$ are the particle and hole states.
The explicit expressions of the single particle matrix elements are obtained
in Appendix \ref{sec:spmel}. This derivation shows that
only unnatural parity states are excited.
This selectivity in exciting the nucleus is a specific feature of the axion
which is a purely pseudo-scalar particle. 
All the other probes excite both natural and unnatural parity states.  

Equation~(\ref{eq:RPAmatel}) shows a sum on the $p$ and $h$ indices indicating 
that the full configuration space spanned by the single-particle states has to be considered.
Usually, this configuration space is obtained by choosing boundary conditions
for all the states, implying that only discrete states are considered.
The sum in $p$ is running up to infinity and it is
obviously truncated in practical calculations. This is the procedure commonly adopted
in the so-called discrete RPA calculations, working well for the description of the 
first nuclear states in the discrete spectrum. 

On the other hand, we are interested in a broad range of axion energies, from few MeV up
to hundreds of MeV. For this kind of excitation spectrum, the discrete RPA is not
adequate since one has to consider also the possibility of a nucleon emission. 
For this reason, we carried out our calculations by considering the CRPA which
can be formally expressed by including in Eq.~(\ref{eq:RPAmatel}) an integration
on the energy of the particle state. The solution of the CRPA equations is not
straightforward as in the discrete RPA case. We briefly present in Appendix \ref{sec:CRPA}
our method of solving the CRPA equations proposed in  Ref.~\cite{don11a}.
Our formulation allows us to use finite-range nucleon-nucleon interactions.
In the CRPA calculations we use the same interaction adopted in the HF calculations
but without Coulomb and spin-orbit terms whose effects are, however, negligible in RPA 
\cite{don14a}.
This CRPA approach has been used to study electron scattering quasi-elastic responses
\cite{don11b} and its extension, treating charge-exchange excitations
\cite{don16}, has been exploited to study neutrino--nucleus scattering processes \cite{co16}.

\section{Axion absorption by $^{16}$O} 
\label{sec:results1}
In this Section, we present the results obtained by applying the theoretical formalism discussed in the previous sections to a specific case: the total absorption of an axion by the $^{16}$O nucleus, the most abundant target for axions in water {\v C}herenkov detectors.

We have already pointed out that we use the same effective nucleon-nucleon interaction
in both the HF and RPA calculations.
We considered
finite-range interactions to obtain the stability of the RPA results without any further
renormalization constant related to the dimensions of the configuration space. The interactions
we have adopted are of Gogny type~\cite{dec80}. More specifically,
we have considered the D1S parametrization~\cite{ber91}
used in the compilation of the AMEDEE data base~\cite{cea,hil07} 
Another parametrization we have considered is the D1M force~\cite{gor09}, fitting
a wider set of data with respect to that used to define the D1S and generating
an equation of state of neutron matter which does not collapse at high densities. 
The last interaction we have used is the D1MT which is obtained by adding to the 
D1M parametrization one tensor and one tensor-isospin term. 
The method used to select the values of the parameters
defining these tensor terms is described in Ref.~\cite{ang11}. The comparison among
the results obtained with the different interaction gives an estimation of the
theoretical uncertainties of our calculations.

\subsection{The ground state}
\begin{figure}[t!] 
\begin{center} 
\includegraphics [width=  0.45\columnwidth,angle=90]{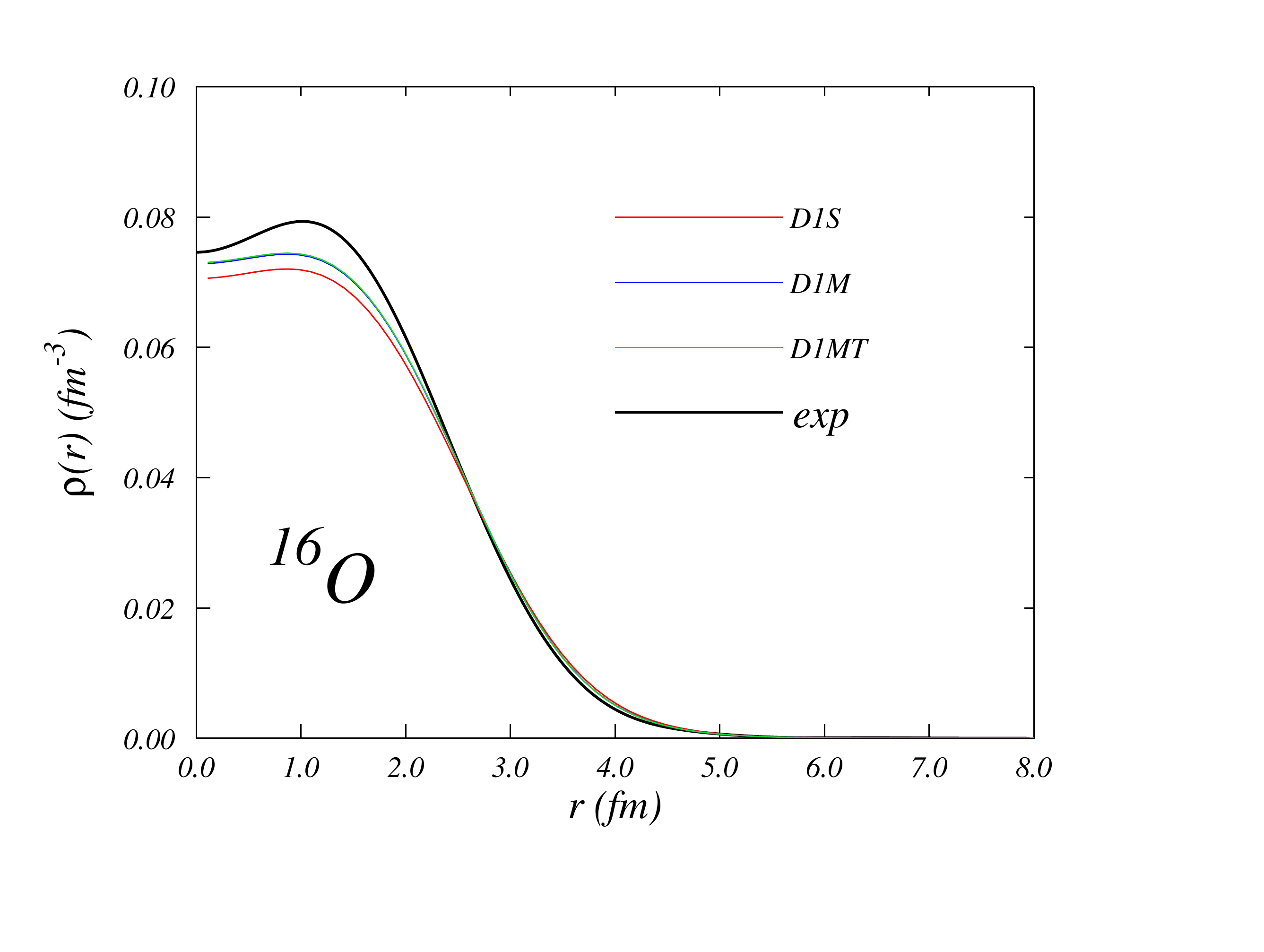} 
\caption{\small Charge densities obtained with the three interactions used in our
work compared with the empirical one (black line) taken from the compilation of
Ref.~\cite{dej87}.
}
\label{fig:dens} 
\end{center} 
\end{figure}

A first validity test of our calculations is done by comparing our results with the observables related to the ground state of $^{16}$O. 
We compare in Tab.~\ref{tab:be} binding energies and root-mean-square (rms) charge radii obtained in our HF calculations with the experimental values taken 
from Ref.~\cite{co85} and Ref.~\cite{led78}, respectively. 
The charge distributions are shown in Fig.~\ref{fig:dens} and the good description
of the surface part of the empirical charge distribution of  Ref.~\cite{dej87}
confirms the agreement with the charge radii shown in Tab.~\ref{tab:be}.

The quality of the description of the ground-state properties obtained with 
the three forces is comparable. Also the single-particle energies of the bound
states shown in Tab.~\ref{tab:spe} are rather similar for the three calculations.
We observe that the energies of the $1p_{1/2}$ states, for both protons and neutrons,
are very similar. The empirical data, taken from Ref.~\cite{co85}, have been
obtained by considering the level scheme of neighboring nuclei. This implies the
use of the so-called Koopmans' theorem which is strictly valid
only in a mean-field description of the nucleus~\cite{rin80}. 
While the empirical single-particle energies of the states below the Fermi level
are reasonably well described by our calculations, this is not the case for those
states above it. In particular, we remark that empirical data show a bound proton $2s_{1/2}$
state, while all our calculations predict it in the continuum.

This good description of the $^{16}$O ground-state is not unexpected, since the
force parameters have been chosen to describe it. The parameter choice is done 
by fitting with HF and Hartree-Fock-Bogolioubov calculations a large set of binding
energies and charge radii all through the set of isotopes composing the nuclear chart, from $Z=6$ up to $Z=130$. This large data set includes also binding energy
and charge radius of $^{16}$O, in which we are interested.

%
% table binding energies
%
\begin{table}[t!]
\begin{center}
\begin{tabular}{l cccc}
\hline
 & D1S & D1M & D1MTd  & exp\\
$B/A$ (MeV) & 7.976 & 8.112 & 7.975 & 8.025 \\
$R_{\rm ch}$ (fm)  & 2.699 & 2.793  & 2.760 & 2.759 \\
\hline
\end{tabular}
\caption{Binding energies per nucleon, $B/A$, and rms charge radii, $R_{\rm ch}$,
obtained in the HF calculations with the interactions used in our work. 
The experimental value of the binding energy is taken from the compilation of Ref.~\cite{bnlw} 
and that of the empirical radius from~\cite{ang13}.}
\label{tab:be}
\end{center}
\end{table}

\begin{table}
\begin{center}
\begin{tabular}{l cccc}
\hline \hline
PROTONS & & & & \\
\hline
                  & D1S & D1M & D1MTd  & exp\\
$2s_{1/2}$ & 1.11 & 0.85 & 0.83 & -0.10 \\
$1d_{5/2}$ & -2.24 & -2.33 & -2.25 & -0.60 \\
\hline
$1p_{1/2}$  & -12.52 & -11.95 & -12.04 & -12.11 \\
$1p_{3/2}$  & -18.62 & -17.67 & -17.74 & -18.44 \\
$1s_{1/2}$  & -35.41 &  -32.78 & -32.93 &  -  \\
\hline\hline
NEUTRONS & & & & \\
\hline
                  & D1S & D1M & D1MTd  & exp\\
$2s_{1/2}$ & -1.87 & -2.21 & -2.23 & -3.27\\
$1d_{5/2}$ & -5.61 & -5.77 & -5.78 & -4.14 \\
\hline
$1p_{1/2}$  & -15.66 & -15.14 & -15.25 & -15.65 \\
$1p_{3/2}$  & -21.86 & -20.95 & -21.02 & -21.81 \\
$1s_{1/2}$  & -38.65 &  -36.09 & -36.19 &  -  \\
\hline\hline
\end{tabular}
\caption{
Energies, in MeV, for bound single-particle states.
The lines divide the particle states, above the Fermi level, from the hole states, 
below the Fermi level. The empirical values, taken from~\cite{co85},
have been obtained from the level scheme of neighbouring nuclei~\cite{led78}.
}
\label{tab:spe}
\end{center}
\end{table}

\subsection{The excitation spectrum}
The description of the excitation spectrum is not as good as the one of the ground state.
Furthermore, the different interactions produce results that are not as uniform as in the 
the ground state case. 
A discussion about the spectrum of the natural parity excitations is done in Ref.~\cite{don11a}.
In this article, we focus our attention on the spectrum of the unnatural parity states since they are
the only ones that axions can excite.

\begin{figure}[t!] 
\begin{center} 
\includegraphics[width=  0.45\columnwidth,angle=90]{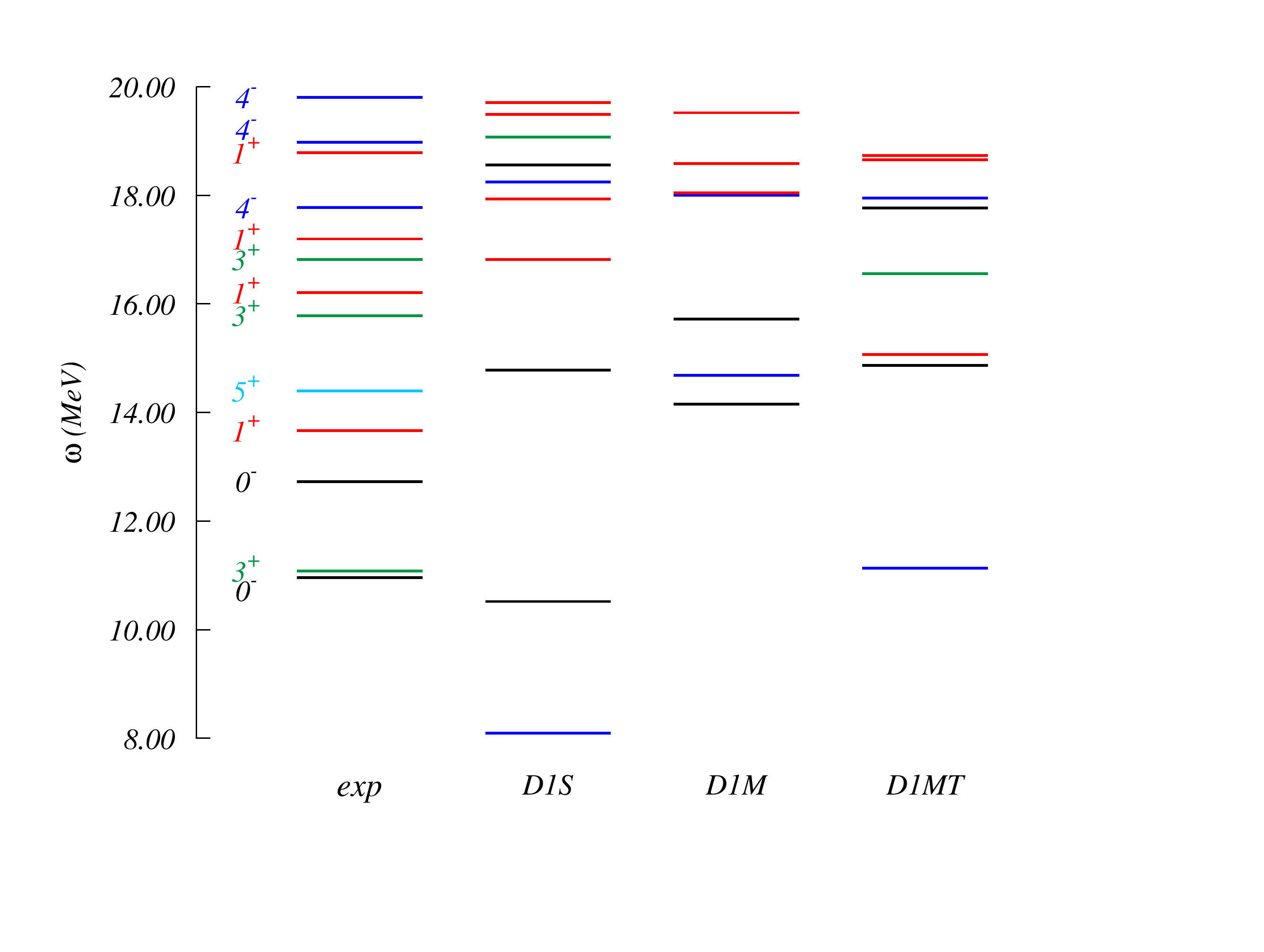} 
\caption{\small 
Spectrum of magnetic states of $^{16}$O. 
Each multipole transition is identified by a specific color:
$0^-$ black, $1^+$ red, $3^+$ green, $4^-$ blue, $5^+$ cyan.
The $2^-$ states are not included.
The experimental energies are taken from the compilation of Ref.~\cite{bnlw}.
}
\label{fig:spec} 
\end{center} 
\end{figure}

In Fig.~\ref{fig:spec} we compare the spectrum of unnatural parity states obtained in our 
discrete RPA calculations with the experimental spectrum taken from Ref.~\cite{bnlw}.
We have considered all the unnatural parity multipoles from $0^-$ up to $5^+$. 
In Fig.~\ref{fig:spec}  we did not insert the $2^-$ states since they are presented separately.
Each multipole is identified by a specific color: black for $0^-$, red for $1^+$, green 
for $3^+$, blue for $4^-$ and cyan for $5^+$. 

The sequence of the levels predicted by our calculations is different from the experimental
one. In addition, each interaction generates a specific sequence of levels. Since we are dealing 
with a nucleus with the same number of protons and neutrons, the type of 
particle-hole $p$-$h$ excitations is the same in both cases. 
For this reason, the identification of excitations of isoscalar (IS) and isovector (IV) type
is relatively easy. In the first case, the $X$ amplitudes of the RPA for protons and neutrons 
have the same sign, and in the IV case they have opposite sign.

The first couple of excited states in the experimental spectrum is that of the $0^-$. 
We observe that an analogous pair of excited states appears in the excitation spectrum 
of the D1S interaction. These two  $0^-$ states are dominated by the 
$(1p_{1/2}^{-1} \otimes 2s_{1/2})$ excitations. The IS state at 10.52 MeV is about 3 MeV
lower than the single-particle excitation energies, while the IV state at 14.78 MeV is roughly
1 MeV higher. This indicates that the interaction is attractive in the IS channel and repulsive 
in the IV one. This picture is not reproduced in the spectrum of the other two  interactions which
are so strong in the IS channel that the excitation energy is below zero. In our calculations 
this produces imaginary solutions of the RPA equations. For these two interactions the first 
$0^-$ states at about 14 MeV are of IV type. 

Another test case is the $4^-$ state. The relatively high value of the angular momentum 
and the negative parity implies that this state is characterized by a simple $p$-$h$ structure.
The lowest energy states are dominated by the $(1p_{3/2}^{-1} \otimes 1d_{5/2})$ 
proton and neutron pairs. Also in this case, we obtain a couple of two excited states 
of IS and IV type whose energies are respectively lower and higher than the energies obtained
by the pure single-particle transitions. 

\begin{figure}[t!] 
\begin{center} 
\includegraphics [ width=0.6\columnwidth,angle=0]{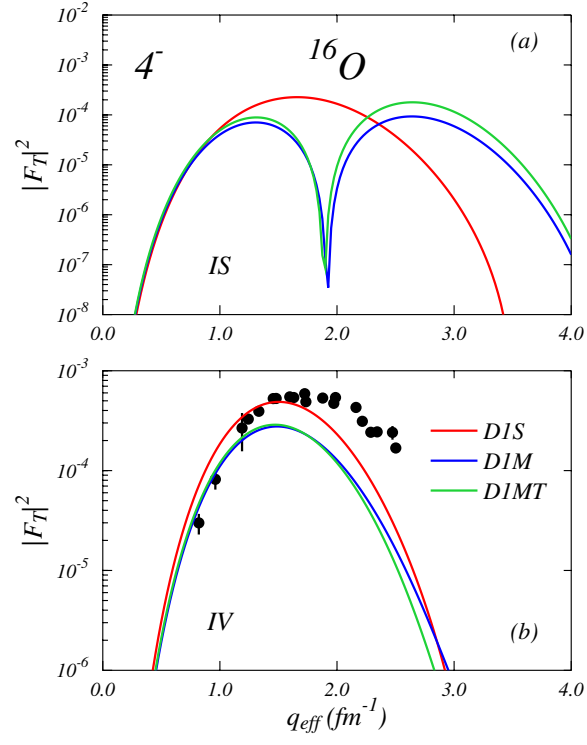} 
\caption{\small 
Electron scattering transverse form factors for the excitation of the $4^-$
excited states. In the upper panel we show the results for the IS state, while
in the lower panel for the IV state.
The experimental data, taken from Ref.~\cite{hyd87}, are those measured at the excitation energy $\omega= 18.975$ MeV. 
}
\label{fig:eem4} 
\end{center} 
\end{figure}

The consequences of the different structure of the RPA solutions for these states are
well pointed out in Fig.~\ref{fig:eem4} where we show the transverse form factors
of an inelastic electron scattering process. The shape of the IV solutions, shown in the 
lower panel of Fig.~\ref{fig:eem4} and compared with the experimental data of Ref.~\cite{hyd87},
are rather similar for all the interactions. On the contrary, the IS solutions (upper panel) obtained
with the D1M and D1MT interactions show a minimum which is not present in the D1S
result. This happens since the D1S result is a pure IS combination of the main $p$-$h$ excitation, i.e.
 $(1p_{3/2}^{-1} \otimes 1d_{5/2})$, as the similarity with the IV case indicates. 
The strong attraction in the IS channel of the D1M and D1MT interaction lowers the pure IS state below zero. In these two cases, the first $4^-$ states contain a strong mixing of the dominant
$(1p_{3/2}^{-1} \otimes 1d_{5/2})$ pair with the $(1p_{3/2}^{-1} \otimes 2d_{5/2})$  and
$(1p_{3/2}^{-1} \otimes 3d_{5/2})$ $p-h$ pairs. The $2d_{5/2}$
 and also $3d_{5/2}$ particle states have additional nodes with respect to the $1d_{5/2}$
 state and this produces the minimum in the form factor.

\begin{figure}[t!] 
\begin{center} 
\includegraphics[width=  0.45\columnwidth,angle=90]{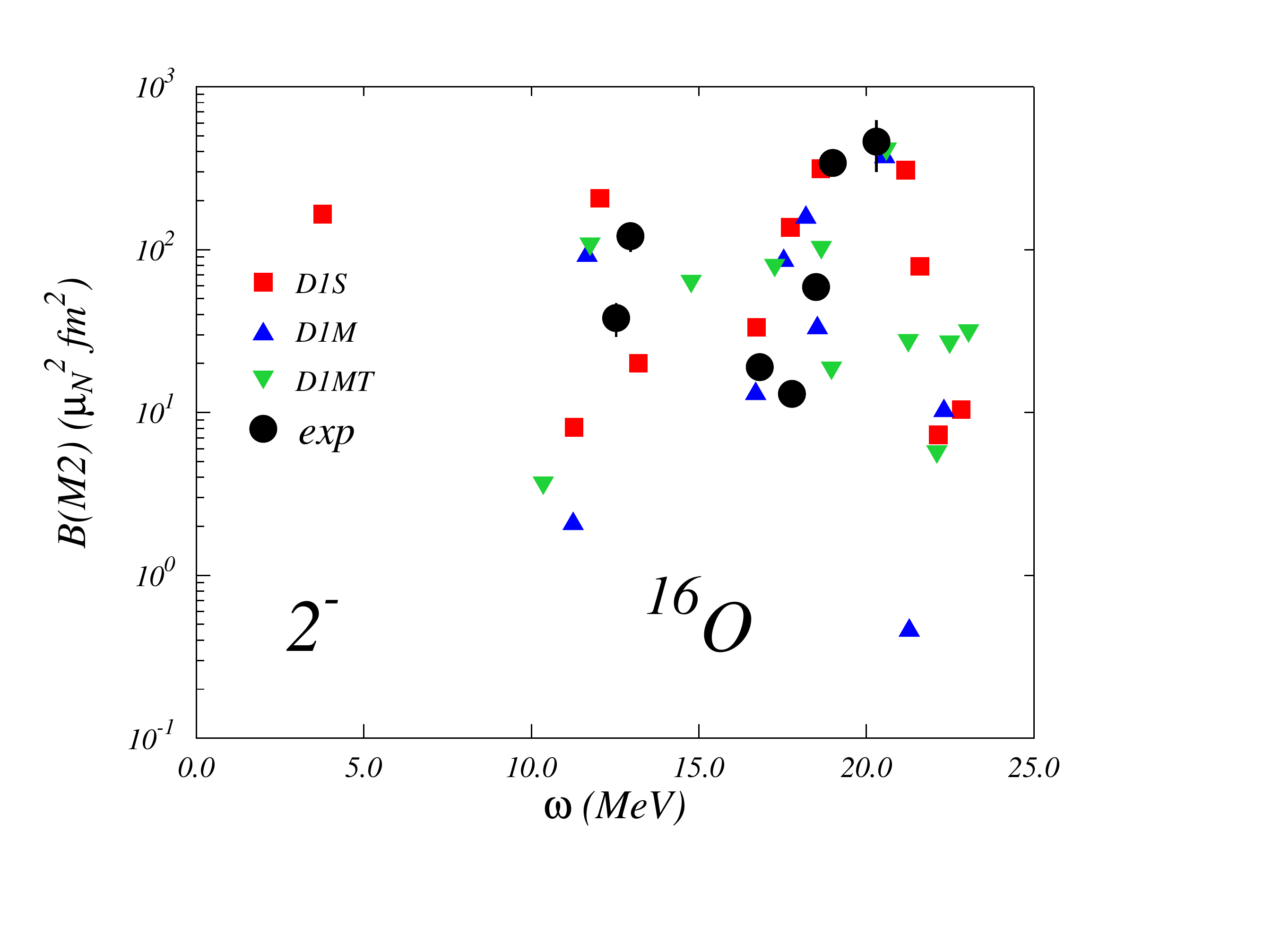} 
\caption{\small 
The $2^-$ excited states of $^{16}$O obtained with our discrete RPA calculations
by using the three interactions we have considered. In this figure we show 
the $B(M2)$ values expressed in terms of $\mu_N^2 {\rm fm}^2$, against the excitation energies. 
The experimental data are taken from Ref.~\cite{kue83}. 
}
\label{fig:m2} 
\end{center} 
\end{figure}

Since the $2^-$ excitation is particularly relevant for the absorption of the axion,
we show in Fig.~\ref{fig:m2} the probability of electromagnetic transition
 $B(M2)$ against the excitation 
energies. We compare our results with the  experimental data taken from Ref.~\cite{kue83}. 
Our calculations generate more states than those experimentally identified. 
The characteristics outlined above for the other excitation multipoles are 
repeated also in this case. 
 In the D1S case, we identify in the states at 3.77 MeV and at 12.05 MeV IS and IV structures
of the  $(1p_{3/2}^{-1} \otimes 1d_{5/2})$ and    $(1p_{1/2}^{-1} \otimes 1d_{5/2})$
protons and neutron excitation pairs. The D1M and D1MT interactions are so strong in the IS
channel that we obtain imaginary solutions of the RPA equations. Also the states
around 12 MeV are a collective mixture of $p$-$h$ excitation pairs which contaminate
the IV character of the main $p$-$h$ pairs.

\begin{figure}[t!] 
\begin{center} 
\includegraphics[width=  0.6\columnwidth,angle=0]{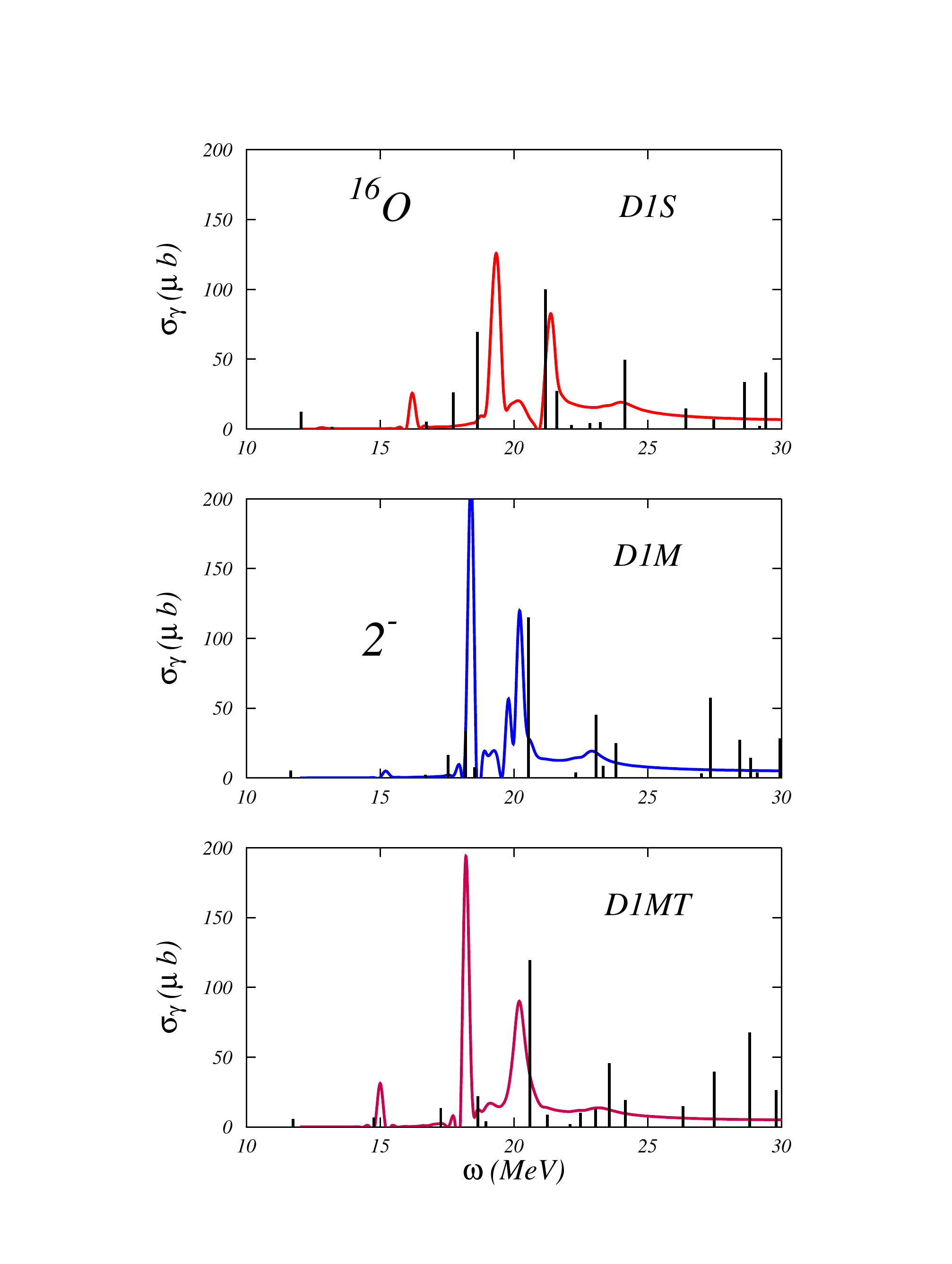} 
\caption{\small 
Total photo-absorption cross section for the excitation of the $2^-$ multipole. 
The curves indicate the CRPA results while the vertical lines the discrete RPA
results. 
}
\label{fig:photom2} 
\end{center} 
\end{figure}

The detailed structure of the excitation spectrum is well understood by 
analysing the RPA solutions. On the other hand, for excitation energies
above 13 MeV the particle emission channel is open. This process is 
described by the CRPA. We compare in Fig.~\ref{fig:photom2} the solution obtained with RPA and 
CRPA calculation for the $2^-$ excitation, by showing the total photo-absorption cross sections calculated by using
the two different theories. The energy of the RPA peaks is
rather similar to that of the CRPA results. In both kinds of calculation
there is a remarkable peak above 20 MeV and another one just below. 
This latter peak is present in all the CRPA calculations but absent in the 
D1MT results obtained with the RPA. 
These results are analogous to those observed for the case of 
natural parity excitation presented in Ref.~\cite{don11a}. It is not possible to compare these results with experimental cross sections
since the contribution of the $2^-$ multipole, which is the largest among 
the unnatural parity states, is three orders of magnitude smaller
than the data~\cite{ahr75}. The total photo-absorption cross section is
dominated by the electric dipole~\cite{don11a}.

\subsection{The axion-nucleus cross section}
\begin{figure}[t!] 
\begin{center} 
\includegraphics [width=  0.45\columnwidth,angle=90]{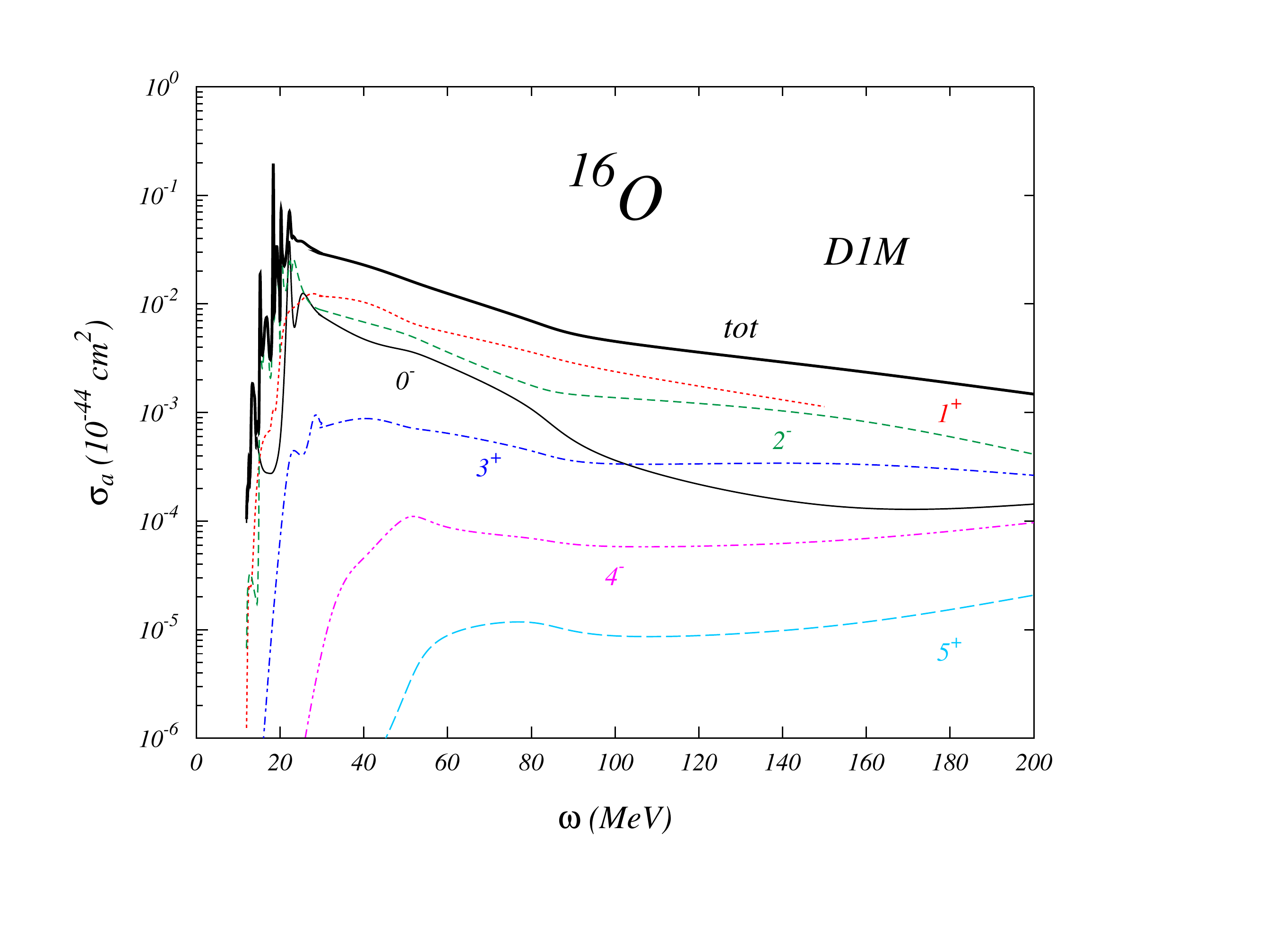} 
\caption{\small 
Total cross section for the absorption of an axion by the $^{16}$O nucleus.
These results have been obtained by using the D1M interaction. The contribution
of the various multipoles is indicated. The thick black line shows the total cross section,
i.e. the sum of all the contributions. 
These results are obtained by using $g_{aN}=2\times10^{-9}$. 
}
\label{fig:axlog} 
\end{center} 
\end{figure}

The CRPA wave functions have been used to calculate the total cross section for the absorption of an axion by the nucleus of $^{16}$O. In all the calculations we used 
the KSVZ-like couplings, i.e. $C_p=-0.47$ and $C_n=0.0$ with
$g_{aN} = 2.0\times10^{-9}$.
This cross section is shown in Fig.~\ref{fig:axlog}  
for the case of the D1M interaction.
The contribution of all the excitation 
multipoles is explicitly shown as well as the total cross section which is the sum
of all these contributions. We have considered energies up to 200 MeV since 
we wanted to study the relevance of the various multipoles at different energy 
regimes. In the range below 20 MeV the cross section is dominated by the $2^-$ excitation. 
This is the reason why, previously, 
we have paid particular attention to this excitation multipole. 
The $0^-$ and $1^+$ multipoles are relatively important at lower energies. 
The contribution of higher multipoles is irrelevant. 

\begin{figure}[t!] 
\begin{center} 
\includegraphics [width=  0.6\columnwidth,angle=0]{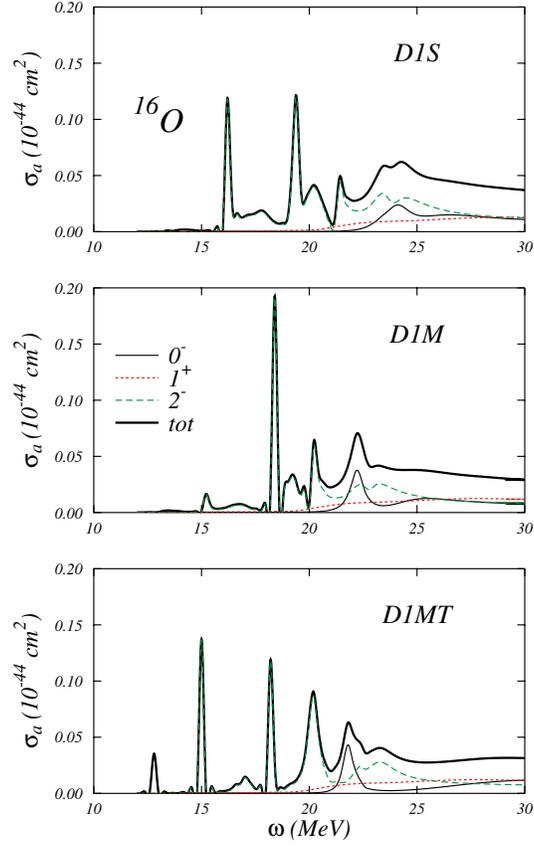} 
\caption{\small 
Total cross section for the absorption of an axion by the $^{16}$O nucleus
in the energy range up to 30 MeV, calculated with
$g_{aN}=2\times10^{-9}$. The three panels show 
the results obtained by using the three interactions we have considered. 
We show separately the contributions of the $0^-$, thin continuous black lines, $1^+$,
dotted red lines and $2^-$, dashed green lines, contributions. 
The thick continuous black lines show the total cross sections total cross section
obtained by summing the contributions of the unnatural parity multipoles up
to $5^+$.
}
\label{fig:xsaxion} 
\end{center} 
\end{figure}

The total axion absorption cross sections up to 30 MeV,
obtained with the three different interactions, are shown
in Fig.~\ref{fig:xsaxion}. The total cross sections have been calculated
by considering all the multipole excitations up to $5^+$, but in the figure
we explicitly present only the 
contribution of the $0^-$, $1^+$ and $2^-$ excitations. 
In all the cases, the $2^-$ cross section coincides with the total one for energies below 20 MeV.
A peak of the $0^-$ at 22-23 MeV is important.
In order to estimate uncertainties related to our nuclear model, we have calculated the total strength of the cross sections, i.e. the energy-integrated cross section, 
by using three different nucleon-nucleon interactions.  The maximum relative difference in these quantities is between the D1S and D1MT results and it amounts to about $4\%$. However, we remark that the difference between the peak values of the cross sections are, 
at most, of the order of $60\%$ because the RPA implicitly gives large uncertainties on the excited levels.

\section{Modelling the radiative decay}
\label{sec:gammad}

For the calculation of the axion induced emission spectra a two-step approach was adopted, similarly to Refs.~\cite{Engel:1990zd,Langanke:1995he}. In a first step, the population of excited states in $^{16}$O by axions was calculated in a RPA approach as explained above. In the second step the de-excitation of the excited state was treated in the subsequent calculation described below.
Excited oxygen states decay via gamma-emission. For states above the particle separation energies, the decay is dominated by particle emission. Emissions of neutrons, protons, $\alpha$-particles, and photons were considered. Due to the fact that highly excited states in $^{16}$O are populated by the interaction with axions, secondary nuclides created after emission of a first particle may also still be in excited states above their particle separation energies. This requires to include the possibility of an emission of another particle from these secondary nuclides. At the initial excitation energies relevant here, it is not necessary to include tertiary particle emission because the tertiary nuclides are created at excitations below their particle emission thresholds. To compute the total gamma-spectra, de-excitations by gamma-cascades in the primary ($^{16}$O), secondary and tertiary nuclides were taken into account. Neutron-, proton- and $\alpha$-emission from $^{16}$O and the secondary nuclides was included in the calculation of the total particle spectra. The summed spectra were obtained with the SPECTRUM code \cite{Rauscher:SPECTRUM} that follows multiple particle emission and gamma-cascades by using pre-calculated transmission coefficients and cascade information, as well as the initial population of the excited states obtained by the RPA.

The transmission coefficients for the four channels were computed using the methods implemented in the SMARAGD Hauser-Feshbach reaction code~\cite{smargd,Rauscher:intjmodphyse}, with spin/parity-selection rules applied in the calculation of the energetically allowed transitions. Particle emission was treated in an optical model, using the microscopic optical potential shown in Ref.~\cite{Jeukenne:1974zz,Lejeune:1980zz} for neutrons and protons. These are especially suited to describe particle emission across the relevant particle-energy range down to low energies as appearing in the secondary emissions. The global optical potential of Ref.~\cite{McFadden:1966zz} was used for $\alpha$-particles, which is specifically also modelled on scattering data for $^{16}$O. The photon emission from an excited state was treated as described in Refs.~\cite{Rauscher:2000fx,Rauscher:intjmodphyse}. Further details on the calculation of the transmission coefficients are found in Ref.~\cite{Rauscher:intjmodphyse}. These transmissions are also used in the SMARAGD code to generate the information required to follow the gamma-cascades and finally to generate the resulting total gamma-spectra from de-excitation of all participating nuclei with the SPECTRUM code. The spectra obtained in this way were then used to predict events in Kamiokande-II, as discussed in the following section.

\section{Expected axion events in Kamiokande-II}
\label{sec:results2}
We used our cross section to revisit the estimation of the number of events 
predicted in the KII water {\v C}herenkov detector, which collected relevant data for neutrino events at the time of SN 1987A~\cite{Hirata:1988ad,Kamiokande-II:1987idp}. 
The number of expected events in the detector can be obtained as~\cite{Fischer:2016cyd}
\begin{equation}
    N_{\mathrm{ev}}=F_a\otimes\,\sigma\,\otimes\,\mathcal{R}\,\otimes\,\mathcal{E}\,,
\label{eq:eventsNumber}
\end{equation}
where $F_a$ is the predicted axion flux from SN 1987A, folded with the detection cross section $\sigma$, the detector energy resolution $\mathcal{R}$ and the detector efficiency $\mathcal{E}$. Notice that the estimation in Eq.~\eqref{eq:eventsNumber} improves the results in Ref.~\cite{Engel:1990zd} by taking into account the detector response to gamma-cascades which occur during the oxygen de-excitation. From Eq.~\eqref{eq:eventsNumber}, the differential events number ${dN_{\mathrm{ev}}}/{dE}$ can be written as~\cite{Lunardini:2004bj}
\begin{equation} 
    \frac{dN_{\mathrm{ev}}}{dE}=N_{\mathrm{T}} \int_0^{+\infty} d\epsilon\,\mathcal{R}_\epsilon(E)\,\mathcal{E}(E)\,F_a(\epsilon)\,\sigma(\epsilon)\,,
\end{equation}
where $\epsilon$ and $E$ are the energy of the incident axion and the detected energy of the de-excitation photon, respectively; $N_\mathrm{T}$ is the number of targets in the detector given by $N_T=N_{\mathrm{A}}\,M/M_{\mathrm{mol}}$, where $N_{\mathrm{A}}=6.02\times10^{23}$ is the Avogadro's number, $M=2.4\,\kton$ is the KII detector mass and $M_{\mathrm{mol}}=18.1\g$ is the molar mass of the oxygen nucleus. In this expression, the detector resolution $\mathcal{R}$ is defined as~\cite{Fogli:2004ff}
\begin{equation}
    \mathcal{R}_\epsilon(E)=\sum_{E_\gamma(\epsilon)}\frac{1}{\sqrt{2\pi\sigma_\gamma    ^2}}\,e^{-(E-E_\gamma(\epsilon))^2/2\sigma_{\gamma}^2}\,BR(E_\gamma(\epsilon))\,,
\end{equation}
where $E_\gamma(\epsilon)$ is the energy of the photon emitted to relax the oxygen nuclei from an excited state of energy $\epsilon$, while $BR(E_\gamma(\epsilon))$ is the branching ratio associated to this process. We highlight that our approach takes into account also the possibility of de-excitation mechanisms without photon emission. Indeed, the oxygen nuclei could also decay by emitting $\alpha$-particles, protons or neutrons, which are undetectable. In particular, the branching ratio for photon emission from a given excited level accounts for $\sim50\%$ of the total de-excitation processes. Furthermore, here we assume that the detected energies are distributed around the true de-excitation energies with a Gaussian line-shape of width $\sigma_\gamma=\sqrt{0.6\,E_\gamma(\epsilon)/\MeV}$, while the detector efficiency can be parameterized as~\cite{Fiorillo:2023frv} 
\begin{equation}
    \mathcal{E}=\begin{cases}\,0\hspace{2.53cm}x<4\\
    \,\frac{0.932}{\sqrt{1+\left(\frac{34}{12-7x+x2}\right)^2}}\,\,\,\,\,\,x\geq4
    \end{cases}\,,    
\end{equation}
where $x=E/\MeV$. Finally, the isotropic axion flux is given by
\begin{equation}
    F_a(\epsilon)=\frac{1}{4\pi D^2}\frac{dN_a(\epsilon)}{d\epsilon}\,,
\end{equation}
where $D=51.4$~kpc is the distance from SN 1987A and 
the axion energy spectra $dN_a/d\epsilon$ are taken from Ref.~\cite{Lella:2023bfb}.

\begin{figure}[t!] 
\begin{center} 
\includegraphics [width=  0.6\columnwidth,angle=0]{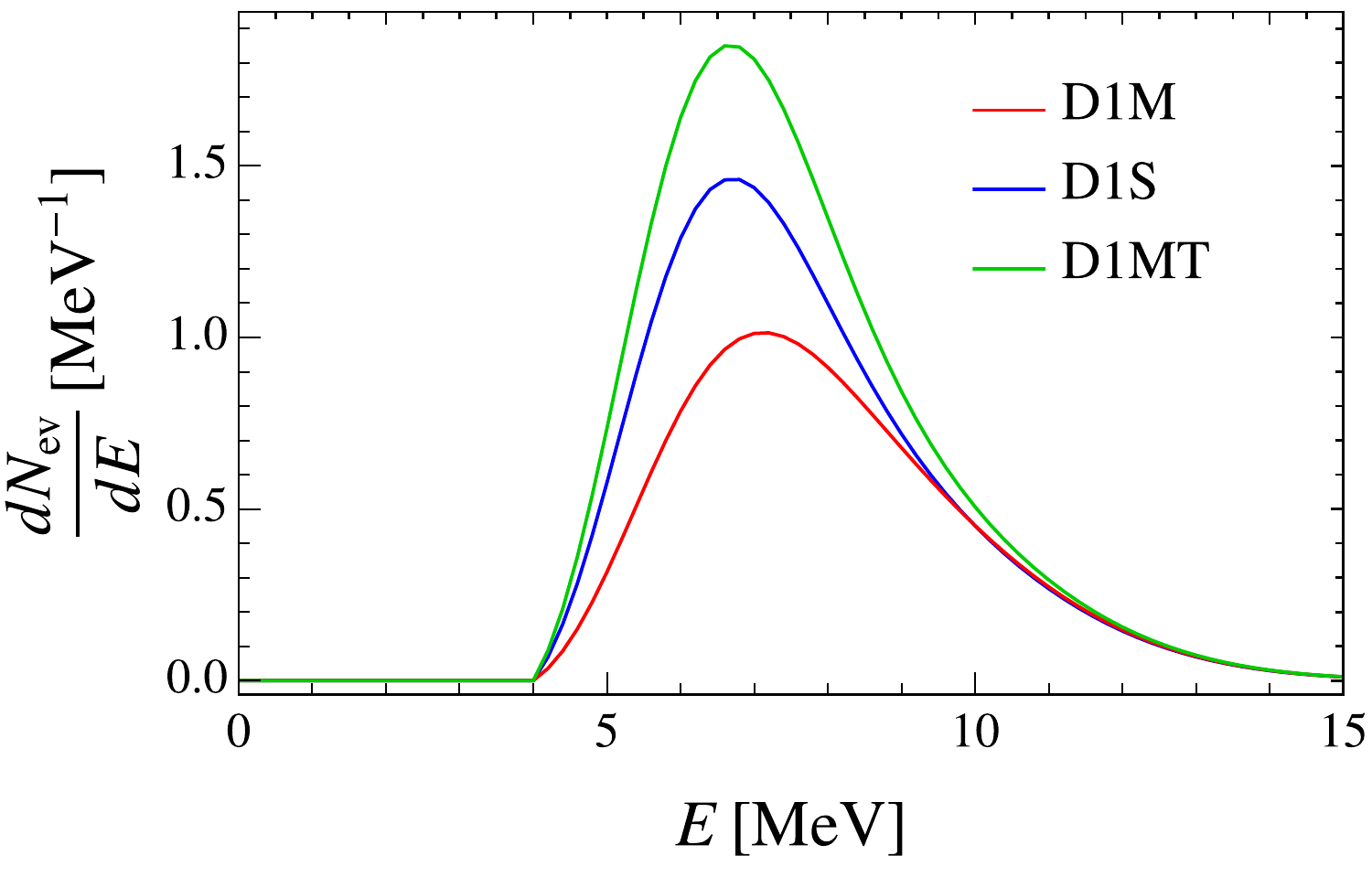} 
\caption{\small 
Differential number of axion-induced events in KII water {\v C}herenkov detector 
for the three different nucleon-nucleon interactions employed. Here the axion-proton coupling is $g_{ap}=10^{-6}$.}
\label{fig:dNevdE} 
\end{center} 
\end{figure}

\begin{figure}[t!] 
\begin{center} 
\includegraphics [width=  0.6\columnwidth,angle=0]{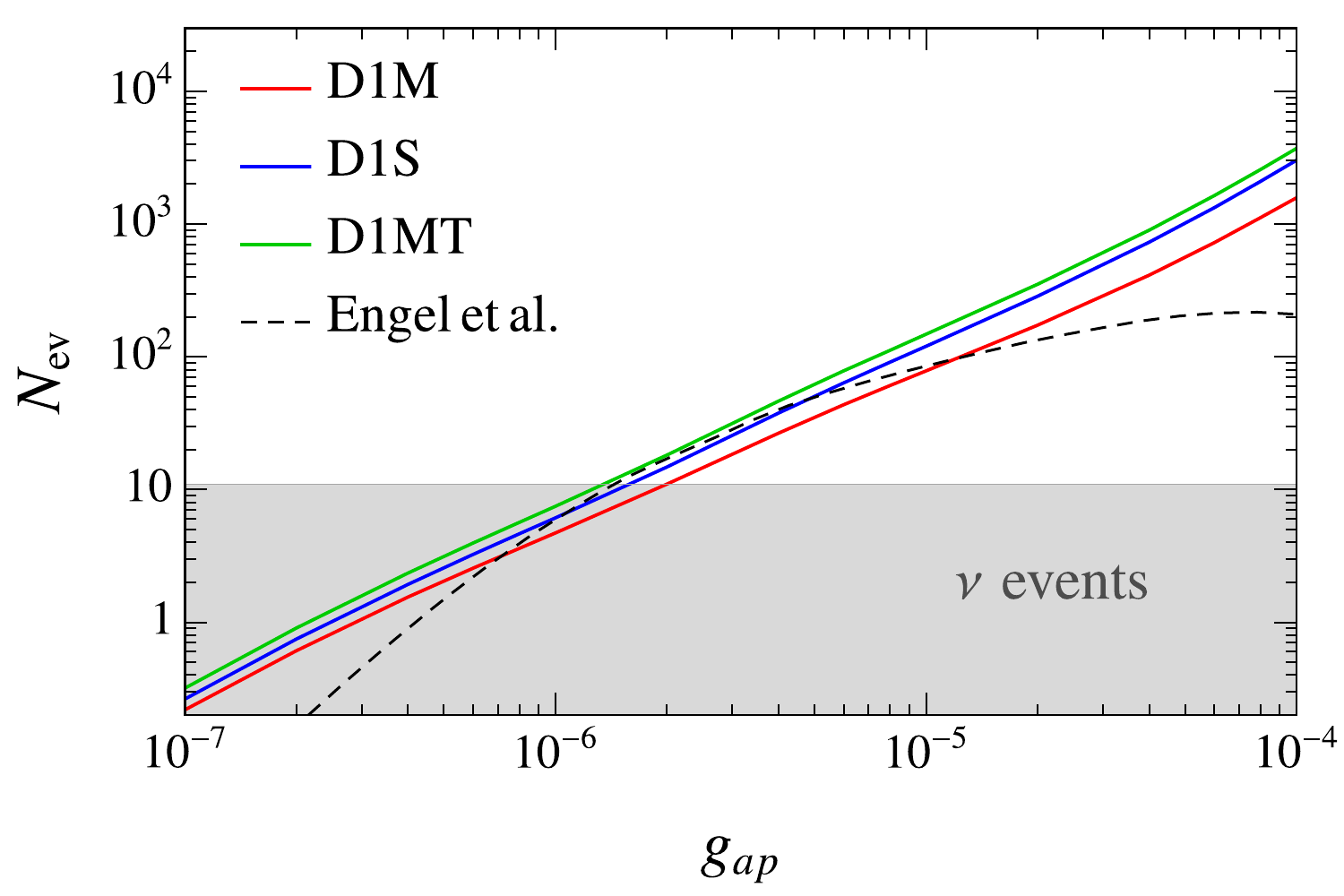} 
\caption{\small 
Predicted number of axion-induced events in KII water {\v C}herenkov detector for the three different nucleon-nucleon interactions employed in the CRPA calculations as a function of the axion-proton coupling. The solid lines are obtained by exploiting the results for cross section introduced in this work and the spectra computed in Ref.~\cite{Lella:2023bfb}, while the dashed lines displays the number of events estimated in Ref.~\cite{Engel:1990zd}. The grey band defines the region where the number of events due to axions is submerged by neutrino events. }
\label{fig:events} 
\end{center} 
\end{figure}

The behavior of  ${dN_{\mathrm{ev}}}/{dE}$ for $g_{ap}=C_p\,g_{aN}=10^{-6}$ obtained with the three different nucleon-nucleon interactions considered
is shown in Fig.~\ref{fig:dNevdE}. We can observe that the detected energies of the de-excitation photons lie in the range $4-10\MeV$ in quite good agreement with Ref.~\cite{Engel:1990zd}. However, in our case the differential number of axion-induced events peaks around $E\sim6-7\MeV$ since the oxygen nuclei mainly relax by emitting photons of energies between $E\simeq4.4\,\MeV$ and $E\simeq8.2\,\MeV$.
These values are slightly different from the main energy of detected photons $E\simeq5.5\MeV$ suggested in Ref.~\cite{Engel:1990zd}.\newline
By integrating over the detected energies it is possible to estimate the number of observed events $N_{\mathrm{ev}}$ in the detector. Figure~\ref{fig:events} shows $N_{\mathrm{ev}}$ as a function of
the axion-proton coupling $g_{ap}$ for the three different effective nucleon-nucleon interactions employed. By comparing our results with the 
detected neutrino events in KII ($\sim11$)~\cite{Hirata:1988ad,Kamiokande-II:1987idp}, we observe that for $g_{ap}\lesssim 10^{-6}$ axion-induced events would have been submerged by neutrino events. Furthermore, it is remarkable to notice that the updated cross-section calculation
gives an event rate comparable to Ref.~\cite{Engel:1990zd} for $g_{ap}\sim10^{-6}-10^{-5}$, while it is higher for $g_{ap}\gtrsim10^{-5}$. In particular, we observe that the number of events obtained in our work quadratically increases with the coupling while in Ref.~\cite{Engel:1990zd} first it saturates, and then it decreases. 
This behavior is mainly due to the different 
SN flux employed. Indeed, while in the simple toy model adopted in Ref.~\cite{Engel:1990zd} the axion
luminosity rapidly decrease for $g_{ap}\gtrsim10^{-4}$, 
Fig.~1 of Ref.~\cite{Lella:2023bfb} (which 
is obtained by integrating over a realistic SN model) shows that for those values of the coupling the axion luminosity 
approaches a plateau (see also Ref.~\cite{Lucente:2022vuo}) and then the dependence on the 
strength of nuclear interactions is essentially determined by the cross section $\sigma\propto\,g_{ap}^2$. Finally, we note that the estimated number of events  depends on the nucleon-nucleon interaction employed in the calculations. 
The uncertainties related to the different choices of the interaction
are $\mathcal{O}(40\%)$ in the weak coupling regime and they increase up to $\mathcal{O}(80\%)$ for strongly-coupled axions.

\section{Summary and Conclusions}
\label{sec:summary}

In this paper we considered the possibility that 
an axion burst emitted from a supernova leads to an observable signal in large underground neutrino water {\v C}herenkov detectors. 
In particular, we referred to the process
$a+ {}^{16}{\rm O}  \to  {}^{16}{\rm O}^{*}$,
where supernova axions may excite the oxygen nuclei in the water of these detectors. The subsequent de-excitation of 
 ${}^{16}{\rm O}^{*}$ may allow for 
 a direct detection of  axions from a Galactic SN. Motivated by this possibility, we presented 
 an updated calculation of the axion-oxygen interaction cross section  using state-of-the-art nuclear models. 
 These latter are based on  the HF theory describing the ground state, and the CRPA to describe the excited states.

We are aware that both HF and RPA are theories based on assumptions neglecting
some physical effects, whose relevance was estimated to be relatively small. 
The RPA considers only
excitations of $1p-1h$ type (see Sec.~9 of Ref.~\cite{co23} for a discussion on the extensions of the RPA theory). 
Since our calculations employ the total strengths of the excitations, a quantity conserved in both RPA and in its extensions, we consider our predictions extremely reliable. Furthermore,
in Ref.~\cite{Engel:1990zd} a phenomenological reduction of the $g_{aN}$ coupling constant for 
the $1^+$ excitation is used to take care of 
the quenching of the Gamow-Teller resonance. Independently of the validity of this phenomenological approach,
which has been under discussion for a long time \cite{ost92}, its implications do not affect our calculations significantly. In particular, the axion-nucleus cross section is dominated by the $2^-$ excitation in the energy range of interest for SN axions, as it is shown in Fig.~\ref{fig:xsaxion}. 
Finally, from our studies of electroweak excitations of the nuclei, we estimate that 
the quantitative impact of other physical effects not considered in our nuclear model, such as short-range
correlations and two-body currents, is negligible (see, e.g., Ref.~\cite{ang06}).
 
After describing the nuclear excitation, we considered the detectable radiative decay of the excited oxygen nucleus. 
We evaluated the nucleus de-excitation by emissions of photons, neutrons, protons, and $\alpha$-particles. The calculations involved the use of transmission coefficients and cascade information obtained from the SMARAGD Hauser-Feshbach reaction code, as well as the SPECTRUM code for multiple particle emission and gamma cascades. 
We used this information to  revise the estimation of 
axion-induced events in the Kamiokande-II detector in relation to the SN 1987A explosion event. 
Our result has been used to obtain updated constraints on the axion-nucleon coupling, in a region complementary with the supernova cooling bound, towards higher coupling $g_{aN}$, as documented in~\cite{Lella:2023bfb}. 

%%%%%%%%%%%%%%%%%%
%%%%%%%%%%%%%%%%%%
\acknowledgements
This article is based upon work from COST Action COSMIC WISPers CA21106, supported by COST (European Cooperation in Science and Technology).
This work is (partially) supported
by ICSC – Centro Nazionale di Ricerca in High Performance Computing,
 Big Data and Quantum Computing, funded by European Union - NextGenerationEU. The work of A.L.,  G.L. and A.M., 
was partially supported by the research grant number 2017W4HA7S ``NAT-NET: Neutrino and Astroparticle Theory Network'' under the program PRIN 2017 funded by the Italian Ministero dell'Università e della Ricerca (MIUR).
The work of P.C. is supported by the European Research Council under Grant No.~742104 and by the Swedish Research Council (VR) under grants  2018-03641 and 2019-02337. 
T.R. was partially supported by the COST action ChETEC (CA16117). P.C., M.G. and G.L. thank the Galileo Galilei Institute for Theoretical Physics for 
hospitality during the preparation of part of this work.

%---------------------------------------------------------
\appendix
%------------------------------------------------------------
\section{Nucleonic current}
\label{app:nuccurrent}
In this Appendix we obtain the explicit expression of the nucleonic current $\mathcal{J}^k$
defined in Eq.~(\ref{eq:Hamiltonian})
\beq
\mathcal{J}^k(\br,t) \equiv \bar{\psi}_f(\br,t)
\gamma^{k}\gamma^{5}(C_{0}+C_{1}\tau_{3})  \psi_i(\br,t)
\;,
\label{eq:ncurr}
\eeq
where $\psi$ indicates the nucleon wave function and $\gamma^{k}$ are the Dirac matrices. Expanding in plane waves the nucleon wavefunctions, we obtain
\beq
\int d^3 r \, \mathcal{J}^k(\br,t) =
(C_0 + C_1 \tau_3) \sqrt{\frac{E_f + m}{2 E_f}} \sqrt{\frac{E_i + m}{2 E_i}} 
\bar{u}(\bp_f,s_f) \gamma^k \gamma^5 u(\bp_i,s_i) e^{-i(E_f-E_i)t}
\;,
\eeq
where $u$ indicates the usual Dirac spinor, $\bar{u}=\gamma^{0}u^{\dagger}$ and $\tau_3=\diag(1,-1)$
the third Pauli matrix.
In the non-relativistic limit we neglect the lower components of the Dirac spinor
and we obtain
\beq
\bar{u}(\bp_f,s_f) \gamma^k \gamma^5 u(\bp_i,s_i) \longrightarrow
\chi^\dagger_{s_f} \sigma^k \chi_{s_i} 
\;,
\eeq
where $\chi_s$ is the Pauli spinor and $\sigma$ a Pauli matrix. 
The $k$-th component of the nucleonic current can be expressed as
\beq
\mathcal{J}^k(\br,0) = (C_0 + C_1 \tau_3) \sigma^k \delta^3(\br - \br_i)
\;\;,
%\label{eq:nucleonic}
\eeq
where $\br_i$ is the generic position where the $i$-th nucleon can be found.

%%%%%%%%%%%%%%%%%%%%%%%%%%%%%%%%%%%%%%%%%%%
\section{Single-particle matrix elements}
\label{sec:spmel}
We work in a spherical basis and we express the single-particle wave functions as
\begin{equation}
\phi(\br)=R^{t}_{nlj}(r) \sum_{\mu=-l}^l \, \sum_{s=-1/2}^{1/2}
\langle l \mu  \frac{1}{2} s | j m\rangle Y_{l,\mu}(\Omega)\chi_{s}\chi_{t}\,,
\equiv R^{t}_{nlj}(r) \chi_{t} \ket{l,j,m}\,,
\end{equation}
where $t$ is the isospin index, $R^{t}_{nlj}$ is the radial wavefunction, $Y_{l,\mu}$ are the spherical harmonics, $\chi_{s}$ 
and $\chi_{t}$ are the spinors for the third component of the spin and isospin,
$n$ is the principal quantum number, $l,j$ are the orbital and total angular momenta, respectively, and $m$ is the $z$-axis component of the latter. We indicate with $\braket{|}$ the Clebsch-Gordan  coefficients.
 
By using the properties of the spherical Bessel functions $j_{J}$ and of the spherical 
harmonics  we can write
\beqn
\nonumber 
&~& \nabla(j_J(pr) Y_{J,0}(\Omega)) = \\
&~& 
p \left[
\sqrt{\frac{J+1}{2J+1}} j_{J+1}(pr) \mathcal{Y}_{J+1,J,0}(\Omega) +
\sqrt{\frac{J}{2J+1}} j_{J-1}(pr) \mathcal{Y}_{J-1,J,0}(\Omega) 
\right]\,,
\eeqn
where we have defined the vector spherical harmonics as
\beq
\mathcal{Y}_{L,J,M} (\Omega) = \sum_{N,q} \braket{L N 1 q | J M} Y_{L, N} {\bf e}_q\,,
\eeq
with $q=\pm,0$ defined as
\beq
{\bf e}_\pm = \mp \frac{1}{\sqrt{2}} (u_x \pm u_y) \;\;\;;\;\;
{\bf e}_0 = u_z\,,
\eeq
where $u_{i}$ for $i=x,y,z$ are unit vectors in Cartesian coordinates. 

The matrix element of Eq.~(\ref{eq:RPAmatel}) between two generic states $a$ and $b$ can be written as 
\beqn
\nonumber
&~& \braket{a ||{\bsigma}\cdot\nabla_i(j_{J}(pr)Y_{J,0}(\Omega))  || b} = 
\\
\nonumber
&p& \left[ \sqrt{\frac{J+1}{2J+1}} 
\int dr\,r^2  j_{J+1}(pr) R^{*\,t}_{n_a l_a j_a}(r) R^{t}_{n_b l_b j_b}(r) 
\braket{l_a j_a m_a || \bsigma \cdot \mathcal{Y}_{J+1J,0} || l_b j_b m_b  }+
\right. \\
&~& \left. +\sqrt{\frac{J}{2J+1}} 
\int dr\,r^2  j_{J-1}(pr) R^{*\,t}_{n_a l_a j_a}(r) R^{t}_{n_b l_b j_b}(r) 
\braket{l_a j_a m_a || \bsigma \cdot \mathcal{Y}_{J-1,J,M} || l_b j_b m_b  }
\right] \,,
\eeqn
with
\beqn
\nonumber
&~&
\braket{l_a j_a m_a || \bsigma \cdot \mathcal{Y}_{J+1J,0} || l_b j_b m_b  } = \\
\nonumber
&~& (-1)^{J+j_b+3/2} \sqrt{\frac{(2j_a+1)(2j_b+1)}{4 \pi} } \xi(l_a+l_b+J+1) 
\\ &~& 
\left[ \frac{\kappa_a + \kappa_b +J +1} {\sqrt{J+1}} \right]
\threej {j_a} {j_b} {J} {\half}{-\half}{0}\,,
\eeqn
and
\beqn
\nonumber
&~&
\braket{l_a j_a m_a || \bsigma \cdot \mathcal{Y}_{J-1J,0} || l_b j_b m_b  } = \\
\nonumber
&~& (-1)^{J+j_b+3/2} \sqrt{\frac{(2j_a+1)(2j_b+1)}{4 \pi} } \xi(l_a+l_b+J+1) 
\\ &~& 
\left[ \frac{\kappa_a + \kappa_b - J} {\sqrt{J+1}} \right]
\threej {j_a} {j_b} {J} {\half}{-\half}{0}\,,
\eeqn
where $\xi(L)=0$ if $L$ is odd and $\xi(L)=1$ if $L$ is even, and 
\beq
\kappa = (l-j)(2j+1)\,.
\eeq
By interchanging the indices $a$ and $b$ we obtain two results differing only by a 
phase factor, therefore Eq.~(\ref{eq:RPAmatel}) can be expressed as 
\beq
\braket{J^\Pi || T_J || 0^+}^{\rm RPA}
= 
\frac{i}{\epsilon_{p}} \int d^3 r \,
\sum_{p,h} \sum_{k=1}^{A}
\Big[X^{J^\Pi}_{ph} + (-1)^{J+j_p+j_h+1} Y^{J^\Pi}_{ph}  \Big]\braket{h ||   
(C_{0}+C_{1}\tau_{3}(k)){\bsigma}\cdot\nabla_i(j_{J}(pr)Y_{J,0}(\Omega))  || p}
\;\;,
\label{eq:RPAmatel1}
\eeq
where $X$ and $Y$ are real
numbers obtained by solving the secular RPA equations \cite{rin80,co23}.

%%%%%%%%%%%%%%%%%%%%%%%%%%%%%%%%%%%%%%%%%
\section{Continuum RPA}
\label{sec:CRPA}

In this Appendix, we indicate with $\epsilon_h$ and $\epsilon_p$ the 
single particle energies of the hole and particle states, respectively.
If the excitation energy $\omega$ of the system is larger than 
$|\epsilon_h|$, the nucleon lying on this 
state can be emitted and leave the system. The 
RPA approach which explicitly considers the emission of a particle is called
Continuum RPA (CRPA), where continuum  refers to the fact that for $\epsilon_p > 0$
the independent particle model Schr\"odinger equation has a continuous spectrum.  
In this case, the single particle wave function has an asymptotically oscillating behavior. 

The RPA secular equations \cite{rin80} with the continuum can be written as 
\begin{equation}
    \begin{split}
 (\epsilon_p-\epsilon_h-\omega)\, X_{ph}^\nu(\epsilon_p)\,+
\sum_{[p']h'} \, \sumint_{\epsilon_{p'}} \,
\left[ \barv_{p h' h p'}(\epsilon_p,\epsilon_{p'})\, X_{p'h'}^\nu(\epsilon_{p'})\,
   + \, \barv_{p p' h h'}(\epsilon_p,\epsilon_{p'}) \, Y_{p'h'}^\nu(\epsilon_{p'}) 
     \right] \, &= \, 0 \,,
\\
(\epsilon_p-\epsilon_h+\omega) \, Y_{ph}^{\nu}(\epsilon_{p})\,+ \sum_{[p']h'} \, \sumint_{\epsilon_{p'}}\, 
\left[
\barv_{h p p' h'}  (\epsilon_p,\epsilon_{p'})\ Y_{p'h'}^{\nu}(\epsilon_{p}) \,
 + \,  \barv_{h h' p p'} (\epsilon_p,\epsilon_{p'}) \, 
X_{p'h'}^{\nu}(\epsilon_{p'})
   \right] \, &= \, 0
\,.
\label{eq:crpa.crpa2}  
    \end{split}
\end{equation}
where $X$ and $Y$ are the unknowns, 
the symbol $[p]$ indicates the set 
of quantum numbers characterizing the particle state except the energy. The states in the continuum are considered by introducing 
the symbol
\beq 
\sumint_{\epsilon_p} \,
\equiv \, \sum_{\epsilon_{\rm F} \le \epsilon_p \le 0} \, + \,
\int_0^\infty {\rm d} \epsilon_p \,,
\label{eq:crpa.simbol}
\eeq
which sums on the discrete energies and integrates on 
the continuum part of the spectrum.  With $\barv_{abcd}$
we indicate the antisymmetrized matrix element
\beq
\barv_{abcd} = \braket{ab | V | cd} - \braket{ab|V|dc}
\;,
\eeq
where $V$ is the effective nucleon-nucleon interaction. 
Our method of solving the CRPA equations \cite{don11a} consists in reformulating the
secular equations in Eq.~(\ref{eq:crpa.crpa2}) with new unknown
functions which do not have explicit dependence on the continuous
particle energy $\epsilon_p$. These unknown functions are the \emph{channel functions} 
$f$ and $g$ defined as
\beq 
f^\nu_{ph}(\br) \, = \, \sumint_{\epsilon_p} \, 
X^\nu_{ph}(\epsilon_p) \, \phi_p(\br,\epsilon_p) \, ,
\label{eq:f}
\eeq 
and
\beq 
g^\nu_{ph}(\br) \, = \, \sumint_{\epsilon_p} \, 
Y^{\nu*}_{ph}(\epsilon_p) \, \phi_p(\br,\epsilon_p)  \, ,
\label{eq:g}
\eeq 
where  $\phi_p(\br,\epsilon_p)$ are eigenfunctions of the single particle Hamiltonian
$\mathcal{H}$. Thus, also $f$ and $g$ are eigenfunctions of the same Hamiltonian
\beq
\mathcal{H} [f^\nu_{ph}(\br)]=\epsilon_p f^\nu_{ph}(\br)  
\;\;{\rm and}\;\;
\mathcal{H} [g^\nu_{ph}(\br)]=\epsilon_p g^\nu_{ph}(\br)  
\;.
\eeq
We multiply the expressions in Eq.~\eqref{eq:crpa.crpa2} by $ \phi_p(\br,\epsilon_p)$, 
and, by exploiting the 
orthonormality properties of the $\phi_p$s we obtain the following equations \cite{don11a}
\begin{equation}
    \begin{split}
  \mathcal{H} [f_{ph}(\br)] 
 -\, (\epsilon_h\, +\, \omega)\, f_{ph}(\br) &=
-\, {\cal F}_{ph}(\br) +\sum_{\epsilon_p<\epsilon_{\rm F}}  \, \phi_p(\br) \, 
\int {\rm d}^3 r_1 \phi^*_h(\br_1) \, {\cal F}_{ph}(\br_1) \, ,
\\
\mathcal{H} [g_{ph}(\br)]  -
(\epsilon_h\, -\, \omega)\, g_{ph}(\br) &=
-\, {\cal G}_{ph}(\br) +\sum_{\epsilon_p<\epsilon_{\rm F}} \, \phi_p(\br) \, 
\int {\rm d}^3 r_1 \, \phi^*_h(\br_1) \, {\cal G}_{ph}(\br_1) \, ,
\label{eq:geq}      
    \end{split}
\end{equation} 
where we have defined 
\beqn
\nonumber
{\cal F}_{ph}(r) \, &=& \, \sum_{[p'] h'} \, \int {\rm d}^3 \,r_2 V(\br,\br_2)
\Bigg\{ \phi^*_{h'}(\br_2) \ \Bigg[
\, \phi_h(\br) \, f_{p'h'}(\br_2) - f_{p'h'}(\br) \, \phi_h (\br_2) \Bigg]
\\
&+& \, g^*_{p'h'}(\br_2) \, 
\Bigg[ \, \phi_h(\br)\, \phi_{h'}(\br_2)\, -  \phi_{h'}(\br) \, \phi_h(\br_2) \Bigg]  
\Bigg\} 
\, ,
\label{eq:Fcal}
\eeqn
and ${\cal G}_{ph}$ is obtained from the above equation by interchanging
the $f$ and $g$ channel functions.  The last terms of Eq.~\eqref{eq:geq} are the contributions of particle wave
functions $\phi_p$ which are not in the continuum.
These equations are solved by imposing the proper
boundary conditions, which in our case consist in indicating 
the particle-hole excitation pair with $\omega > |\epsilon_h|$ from where the nucleon
is emitted. These particle-hole ($ph$) pairs are called {\it open channels} and the 
CRPA secular equations are solved a number of times equal
to the number of the open channels, each by imposing that  the particle 
is emitted only in one open channel called elastic.
The boundary conditions 
are imposed on the radial parts of the $f$ and $g$ functions.
For an open $ph$ channel, the outgoing
asymptotic behavior of the channel function $f_{ph}^{p_0 h_0}$ is 
\beq
f_{ph}^{p_0 h_0}(r\to\infty)\, \to \,
R_{p_0}(r,\epsilon_p)\, \delta_{p,p_0}\, \delta_{h,h_0}\, +\, \lambda
\, H^-_p(\epsilon_h+\omega,r)
\,,
\label{eq:asintf} 
\eeq 
where $\lambda$ is a complex normalization constant and
$H^-_p(\epsilon_h+\omega,r)$ is an ingoing Coulomb function 
if the emitted particle is electrically charged or a Hankel function for neutrons.
The radial part of the single particle wave function $R_{p}$ is the eigenfunction of $\mathcal{H}$
for positive energy. In the case of a closed channel, the asymptotic behaviour is given by 
a decreasing exponential function 
\beq
f_{ph}^{p_0 h_0}(r\to\infty) \, 
\rightarrow \, 
\frac{1}{r}\,
\exp\left[-r\left(\frac{2m|\epsilon_h+\omega|}{\hbar^2}\right)^{\frac{1}{2}}
\right]
\,,
\label{eq:asintf1}
\eeq
in analogy to the case of the channel functions
$g_{ph}^{p_0 h_0}$, 
\beq
g_{ph}^{p_0 h_0}(r\to\infty)
\, \rightarrow \,
\frac{1}{r} \,
\exp\left[-r\left(\frac{2m|\epsilon_h-\omega|}{\hbar^2}\right)^{\frac{1}{2}}
\right]
\,.
\label{eq:asintg}
\eeq
We solve these CRPA secular equations by expanding $f$ and $g$ 
on the  basis of Sturm functions $\Phi^{\mu}_p$ which obey the required 
boundary conditions in Eq.~(\ref{eq:asintf})--Eq.~(\ref{eq:asintg})~\cite{don11a}.
In the independent particle model, i.e. $V=0$, 
the particle emission process is described by considering
that a particle lying on the hole state $h_0$ is emitted in the particle state $p_0$.
The CRPA considers this fact in the elastic channel and, in addition, 
takes care of the fact that the residual interaction
mixes this direct emission with all the other $ph$ pairs compatible 
with the total angular momentum of the excitation. 
In Appendix \ref{sec:spmel} we have shown that the one-body transition operator 
can be expressed in the form
\beq
T_{J,M}(\br) = \sum_{i=1}^A F_J(r_i) \theta_{J,M}(\Omega_i) \delta^3(\br_i - \br) 
\;,
\eeq
where we have separated the dependence on the radial and angular coordinates in $F_{J}(r_{i})$ and $\theta_{J,M}(\Omega_{i})$, respectively. 
The transition matrix element  in Eq.~(\ref{eq:RPAmatel1}) for a specific open channel 
$p_0 h_0$ is 
\beqn
\nonumber
\braket{J^\Pi || T_J || 0^+}^{\rm RPA}_{p_0 h_0} 
&=& \sum_{ph} \Big[\braket{ j_p || \theta_J || j_h }
\int dr r^2 \left(f^{p_0 h_0}_{ph} (r) \right)^* R_h(r) F_J(r) +
\\
&~&+(-1)^{J+j_p-j_h} \braket{ j_h || \theta_J || j_p }
\int dr r^2 g^{p_0 h_0}_{ph} (r)  R^*_h(r) F_J(r) 
\Big]\;\;.
\label{eq:CRPAmatel}
\eeqn
The cross section is obtained by summing, incoherently, on all the open channels $p_0 h_0$
the square moduli of the above transition amplitude.

%%%%%%%%%%%%%%%%%%%%%%%%%%%%%%%%%%%%%%%%%%%%%%%%%%%%%% 
%   Bibliography 
%%%%%%%%%%%%%%%%%%%%%%%%%%%%%%%%%%%%%%%%%%%%%%%%%%%%%% 
  
%
\bibliographystyle{elsarticle-num}  
\bibliography{biblio.bib}

\begin{thebibliography}{10}
\expandafter\ifx\csname url\endcsname\relax
  \def\url#1{\texttt{#1}}\fi
\expandafter\ifx\csname urlprefix\endcsname\relax\def\urlprefix{URL }\fi
\expandafter\ifx\csname href\endcsname\relax
  \def\href#1#2{#2} \def\path#1{#1}\fi

\bibitem{Raffelt:1999tx}
G.~G. Raffelt, {Particle physics from stars}, Ann. Rev. Nucl. Part. Sci. 49
  (1999) 163--216.
\newblock \href {http://arxiv.org/abs/hep-ph/9903472}
  {\path{arXiv:hep-ph/9903472}}, \href
  {https://doi.org/10.1146/annurev.nucl.49.1.163}
  {\path{doi:10.1146/annurev.nucl.49.1.163}}.

\bibitem{Kamiokande-II:1987idp}
K.~Hirata, et~al., {Observation of a Neutrino Burst from the Supernova SN
  1987a}, Phys. Rev. Lett. 58 (1987) 1490--1493.
\newblock \href {https://doi.org/10.1103/PhysRevLett.58.1490}
  {\path{doi:10.1103/PhysRevLett.58.1490}}.

\bibitem{Hirata:1988ad}
K.~S. Hirata, et~al., {Observation in the Kamiokande-II Detector of the
  Neutrino Burst from Supernova SN 1987a}, Phys. Rev. D 38 (1988) 448--458.
\newblock \href {https://doi.org/10.1103/PhysRevD.38.448}
  {\path{doi:10.1103/PhysRevD.38.448}}.

\bibitem{IMB:1988suc}
C.~B. Bratton, et~al., {Angular Distribution of Events From Sn1987a}, Phys.
  Rev. D 37 (1988) 3361.
\newblock \href {https://doi.org/10.1103/PhysRevD.37.3361}
  {\path{doi:10.1103/PhysRevD.37.3361}}.

\bibitem{Bionta:1987qt}
R.~M. Bionta, et~al., {Observation of a Neutrino Burst in Coincidence with
  Supernova SN 1987a in the Large Magellanic Cloud}, Phys. Rev. Lett. 58 (1987)
  1494.
\newblock \href {https://doi.org/10.1103/PhysRevLett.58.1494}
  {\path{doi:10.1103/PhysRevLett.58.1494}}.

\bibitem{Raffelt:1987yt}
G.~Raffelt, D.~Seckel, {Bounds on Exotic Particle Interactions from SN 1987a},
  Phys. Rev. Lett. 60 (1988) 1793.
\newblock \href {https://doi.org/10.1103/PhysRevLett.60.1793}
  {\path{doi:10.1103/PhysRevLett.60.1793}}.

\bibitem{Raffelt:1990yu}
G.~G. Raffelt, {What Have We Learned From SN 1987A?}, Mod. Phys. Lett. A 5
  (1990) 2581--2592.
\newblock \href {https://doi.org/10.1142/S0217732390003000}
  {\path{doi:10.1142/S0217732390003000}}.

\bibitem{Keil:1996ju}
W.~Keil, H.-T. Janka, D.~N. Schramm, G.~Sigl, M.~S. Turner, J.~R. Ellis, {A
  Fresh look at axions and SN-1987A}, Phys. Rev. D 56 (1997) 2419--2432.
\newblock \href {http://arxiv.org/abs/astro-ph/9612222}
  {\path{arXiv:astro-ph/9612222}}, \href
  {https://doi.org/10.1103/PhysRevD.56.2419}
  {\path{doi:10.1103/PhysRevD.56.2419}}.

\bibitem{Chang:2018rso}
J.~H. Chang, R.~Essig, S.~D. McDermott, {Supernova 1987A Constraints on Sub-GeV
  Dark Sectors, Millicharged Particles, the QCD Axion, and an Axion-like
  Particle}, JHEP 09 (2018) 051.
\newblock \href {http://arxiv.org/abs/1803.00993} {\path{arXiv:1803.00993}},
  \href {https://doi.org/10.1007/JHEP09(2018)051}
  {\path{doi:10.1007/JHEP09(2018)051}}.

\bibitem{Carenza:2019pxu}
P.~Carenza, T.~Fischer, M.~Giannotti, G.~Guo, G.~Mart\'\i{}nez-Pinedo,
  A.~Mirizzi, {Improved axion emissivity from a supernova via nucleon-nucleon
  bremsstrahlung}, JCAP 10~(10) (2019) 016, [Erratum: JCAP 05, E01 (2020)].
\newblock \href {http://arxiv.org/abs/1906.11844} {\path{arXiv:1906.11844}},
  \href {https://doi.org/10.1088/1475-7516/2019/10/016}
  {\path{doi:10.1088/1475-7516/2019/10/016}}.

\bibitem{Carenza:2020cis}
P.~Carenza, B.~Fore, M.~Giannotti, A.~Mirizzi, S.~Reddy, {Enhanced Supernova
  Axion Emission and its Implications}, Phys. Rev. Lett. 126~(7) (2021) 071102.
\newblock \href {http://arxiv.org/abs/2010.02943} {\path{arXiv:2010.02943}},
  \href {https://doi.org/10.1103/PhysRevLett.126.071102}
  {\path{doi:10.1103/PhysRevLett.126.071102}}.

\bibitem{Caputo:2021rux}
A.~Caputo, G.~Raffelt, E.~Vitagliano, {Muonic boson limits: Supernova redux},
  Phys. Rev. D 105~(3) (2022) 035022.
\newblock \href {http://arxiv.org/abs/2109.03244} {\path{arXiv:2109.03244}},
  \href {https://doi.org/10.1103/PhysRevD.105.035022}
  {\path{doi:10.1103/PhysRevD.105.035022}}.

\bibitem{Kolb:1996pa}
E.~W. Kolb, R.~N. Mohapatra, V.~L. Teplitz, {New supernova constraints on
  sterile neutrino production}, Phys. Rev. Lett. 77 (1996) 3066--3069.
\newblock \href {http://arxiv.org/abs/hep-ph/9605350}
  {\path{arXiv:hep-ph/9605350}}, \href
  {https://doi.org/10.1103/PhysRevLett.77.3066}
  {\path{doi:10.1103/PhysRevLett.77.3066}}.

\bibitem{Raffelt:2011nc}
G.~G. Raffelt, S.~Zhou, {Supernova bound on keV-mass sterile neutrinos
  reexamined}, Phys. Rev. D 83 (2011) 093014.
\newblock \href {http://arxiv.org/abs/1102.5124} {\path{arXiv:1102.5124}},
  \href {https://doi.org/10.1103/PhysRevD.83.093014}
  {\path{doi:10.1103/PhysRevD.83.093014}}.

\bibitem{Mastrototaro:2019vug}
L.~Mastrototaro, A.~Mirizzi, P.~D. Serpico, A.~Esmaili, {Heavy sterile neutrino
  emission in core-collapse supernovae: Constraints and signatures}, JCAP 01
  (2020) 010.
\newblock \href {http://arxiv.org/abs/1910.10249} {\path{arXiv:1910.10249}},
  \href {https://doi.org/10.1088/1475-7516/2020/01/010}
  {\path{doi:10.1088/1475-7516/2020/01/010}}.

\bibitem{Chang:2016ntp}
J.~H. Chang, R.~Essig, S.~D. McDermott, {Revisiting Supernova 1987A Constraints
  on Dark Photons}, JHEP 01 (2017) 107.
\newblock \href {http://arxiv.org/abs/1611.03864} {\path{arXiv:1611.03864}},
  \href {https://doi.org/10.1007/JHEP01(2017)107}
  {\path{doi:10.1007/JHEP01(2017)107}}.

\bibitem{Dev:2020eam}
P.~S.~B. Dev, R.~N. Mohapatra, Y.~Zhang, {Revisiting supernova constraints on a
  light CP-even scalar}, JCAP 08 (2020) 003, [Erratum: JCAP 11, E01 (2020)].
\newblock \href {http://arxiv.org/abs/2005.00490} {\path{arXiv:2005.00490}},
  \href {https://doi.org/10.1088/1475-7516/2020/08/003}
  {\path{doi:10.1088/1475-7516/2020/08/003}}.

\bibitem{Camalich:2020wac}
J.~M. Camalich, J.~Terol-Calvo, L.~Tolos, R.~Ziegler, {Supernova Constraints on
  Dark Flavored Sectors}, Phys. Rev. D 103~(12) (2021) L121301.
\newblock \href {http://arxiv.org/abs/2012.11632} {\path{arXiv:2012.11632}},
  \href {https://doi.org/10.1103/PhysRevD.103.L121301}
  {\path{doi:10.1103/PhysRevD.103.L121301}}.

\bibitem{Hannestad:2007ys}
S.~Hannestad, G.~Raffelt, Y.~Y.~Y. Wong, {Unparticle constraints from SN
  1987A}, Phys. Rev. D 76 (2007) 121701.
\newblock \href {http://arxiv.org/abs/0708.1404} {\path{arXiv:0708.1404}},
  \href {https://doi.org/10.1103/PhysRevD.76.121701}
  {\path{doi:10.1103/PhysRevD.76.121701}}.

\bibitem{Mirizzi:316773}
A.~Mirizzi, P.~D. Serpico, G.~Sigl (Eds.),
  \href{https://bib-pubdb1.desy.de/record/316773}{{R}affelt, {G}eorg
  "{S}upernova as particle-physics laboratory" in {H}amburg neutrinos from
  supernova explosions. {P}roceedings, {W}orkshop, {HANSE} 2011}, DESY-PROC,
  Hamburg Neutrinos From Supernova Explosions, Hamburg (Germany), 19 Jul 2011 -
  23 Jul 2011, Verlag Deutsches Elektronen-Synchrotron, Hamburg, 2011.
\newblock \href {https://doi.org/10.3204/DESY-PROC-2011-03/raffelt}
  {\path{doi:10.3204/DESY-PROC-2011-03/raffelt}}.
\newline\urlprefix\url{https://bib-pubdb1.desy.de/record/316773}

\bibitem{Antel:2023hkf}
C.~Antel, et~al., {Feebly Interacting Particles: FIPs 2022 workshop report},
  in: {Workshop on Feebly-Interacting Particles}, 2023.
\newblock \href {http://arxiv.org/abs/2305.01715} {\path{arXiv:2305.01715}}.

\bibitem{Engel:1990zd}
J.~Engel, D.~Seckel, A.~C. Hayes, {Emission and detectability of hadronic
  axions from SN1987A}, Phys. Rev. Lett. 65 (1990) 960--963.
\newblock \href {https://doi.org/10.1103/PhysRevLett.65.960}
  {\path{doi:10.1103/PhysRevLett.65.960}}.

\bibitem{don11a}
V.~{De Donno}, G.~Co', M.~Anguiano, A.~M. Lallena, Self-consistent continuum
  random-phase approximation calculations with finite-range interactions, Phys.
  \ Rev. \ C 83 (2011) 044324.

\bibitem{Kim:1979if}
J.~E. Kim, {Weak Interaction Singlet and Strong CP Invariance}, Phys. Rev.
  Lett. 43 (1979) 103.
\newblock \href {https://doi.org/10.1103/PhysRevLett.43.103}
  {\path{doi:10.1103/PhysRevLett.43.103}}.

\bibitem{Shifman:1979if}
M.~A. Shifman, A.~Vainshtein, V.~I. Zakharov, {Can Confinement Ensure Natural
  CP Invariance of Strong Interactions?}, Nucl. Phys. B 166 (1980) 493--506.
\newblock \href {https://doi.org/10.1016/0550-3213(80)90209-6}
  {\path{doi:10.1016/0550-3213(80)90209-6}}.

\bibitem{DiLuzio:2017ogq}
L.~Di~Luzio, F.~Mescia, E.~Nardi, P.~Panci, R.~Ziegler, {Astrophobic Axions},
  Phys. Rev. Lett. 120~(26) (2018) 261803.
\newblock \href {http://arxiv.org/abs/1712.04940} {\path{arXiv:1712.04940}},
  \href {https://doi.org/10.1103/PhysRevLett.120.261803}
  {\path{doi:10.1103/PhysRevLett.120.261803}}.

\bibitem{DiLuzio:2020wdo}
L.~Di~Luzio, M.~Giannotti, E.~Nardi, L.~Visinelli, {The landscape of QCD axion
  models}, Phys. Rept. 870 (2020) 1--117.
\newblock \href {http://arxiv.org/abs/2003.01100} {\path{arXiv:2003.01100}},
  \href {https://doi.org/10.1016/j.physrep.2020.06.002}
  {\path{doi:10.1016/j.physrep.2020.06.002}}.

\bibitem{DiLuzio:2021ysg}
L.~Di~Luzio, M.~Fedele, M.~Giannotti, F.~Mescia, E.~Nardi, {Stellar evolution
  confronts axion models}, JCAP 02~(02) (2022) 035.
\newblock \href {http://arxiv.org/abs/2109.10368} {\path{arXiv:2109.10368}},
  \href {https://doi.org/10.1088/1475-7516/2022/02/035}
  {\path{doi:10.1088/1475-7516/2022/02/035}}.

\bibitem{Moriyama:1995bz}
S.~Moriyama, {A Proposal to search for a monochromatic component of solar
  axions using Fe-57}, Phys. Rev. Lett. 75 (1995) 3222--3225.
\newblock \href {http://arxiv.org/abs/hep-ph/9504318}
  {\path{arXiv:hep-ph/9504318}}, \href
  {https://doi.org/10.1103/PhysRevLett.75.3222}
  {\path{doi:10.1103/PhysRevLett.75.3222}}.

\bibitem{Krcmar:1998xn}
M.~Krcmar, Z.~Krecak, M.~Stipcevic, A.~Ljubicic, D.~A. Bradley, {Search for
  invisible axions using Fe-57}, Phys. Lett. B 442 (1998) 38.
\newblock \href {http://arxiv.org/abs/nucl-ex/9801005}
  {\path{arXiv:nucl-ex/9801005}}, \href
  {https://doi.org/10.1016/S0370-2693(98)01231-3}
  {\path{doi:10.1016/S0370-2693(98)01231-3}}.

\bibitem{CAST:2009jdc}
S.~Andriamonje, et~al., {Search for 14.4-keV solar axions emitted in the
  M1-transition of Fe-57 nuclei with CAST}, JCAP 12 (2009) 002.
\newblock \href {http://arxiv.org/abs/0906.4488} {\path{arXiv:0906.4488}},
  \href {https://doi.org/10.1088/1475-7516/2009/12/002}
  {\path{doi:10.1088/1475-7516/2009/12/002}}.

\bibitem{CUORE:2012ymr}
F.~Alessandria, et~al., {Search for 14.4 keV solar axions from M1 transition of
  Fe-57 with CUORE crystals}, JCAP 05 (2013) 007.
\newblock \href {http://arxiv.org/abs/1209.2800} {\path{arXiv:1209.2800}},
  \href {https://doi.org/10.1088/1475-7516/2013/05/007}
  {\path{doi:10.1088/1475-7516/2013/05/007}}.

\bibitem{DiLuzio:2021qct}
L.~Di~Luzio, et~al., {Probing the axion\textendash{}nucleon coupling with the
  next generation of~axion helioscopes}, Eur. Phys. J. C 82~(2) (2022) 120.
\newblock \href {http://arxiv.org/abs/2111.06407} {\path{arXiv:2111.06407}},
  \href {https://doi.org/10.1140/epjc/s10052-022-10061-1}
  {\path{doi:10.1140/epjc/s10052-022-10061-1}}.

\bibitem{Borexino:2012guz}
G.~Bellini, et~al., {Search for Solar Axions Produced in $p(d,\rm{^3He})A$
  Reaction with Borexino Detector}, Phys. Rev. D 85 (2012) 092003.
\newblock \href {http://arxiv.org/abs/1203.6258} {\path{arXiv:1203.6258}},
  \href {https://doi.org/10.1103/PhysRevD.85.092003}
  {\path{doi:10.1103/PhysRevD.85.092003}}.

\bibitem{Bhusal:2020bvx}
A.~Bhusal, N.~Houston, T.~Li, {Searching for Solar Axions Using Data from the
  Sudbury Neutrino Observatory}, Phys. Rev. Lett. 126~(9) (2021) 091601.
\newblock \href {http://arxiv.org/abs/2004.02733} {\path{arXiv:2004.02733}},
  \href {https://doi.org/10.1103/PhysRevLett.126.091601}
  {\path{doi:10.1103/PhysRevLett.126.091601}}.

\bibitem{Lucente:2022esm}
G.~Lucente, N.~Nath, F.~Capozzi, M.~Giannotti, A.~Mirizzi, {Probing high-energy
  solar axion flux with a large scintillation neutrino detector}, Phys. Rev. D
  106~(12) (2022) 123007.
\newblock \href {http://arxiv.org/abs/2209.11780} {\path{arXiv:2209.11780}},
  \href {https://doi.org/10.1103/PhysRevD.106.123007}
  {\path{doi:10.1103/PhysRevD.106.123007}}.

\bibitem{Alonso:2021kyu}
J.~Alonso, et~al., {Neutrino physics opportunities with the IsoDAR source at
  Yemilab}, Phys. Rev. D 105~(5) (2022) 052009.
\newblock \href {http://arxiv.org/abs/2111.09480} {\path{arXiv:2111.09480}},
  \href {https://doi.org/10.1103/PhysRevD.105.052009}
  {\path{doi:10.1103/PhysRevD.105.052009}}.

\bibitem{Waites:2022tov}
L.~Waites, A.~Thompson, A.~Bungau, J.~M. Conrad, B.~Dutta, W.-C. Huang, D.~Kim,
  M.~Shaevitz, J.~Spitz, {Axion-Like Particle Production at Beam Dump
  Experiments with Distinct Nuclear Excitation Lines} (7 2022).
\newblock \href {http://arxiv.org/abs/2207.13659} {\path{arXiv:2207.13659}}.

\bibitem{Lella:2023bfb}
A.~Lella, P.~Carenza, G.~Co', G.~Lucente, M.~Giannotti, A.~Mirizzi,
  T.~Rauscher, {Getting the most on supernova axions} (6 2023).
\newblock \href {http://arxiv.org/abs/2306.01048} {\path{arXiv:2306.01048}}.

\bibitem{bjo64}
J.~D. Bjorken, S.~D. Drell, Relativistic Quantum Mechanics, McGraw-Hill, New
  York, 1964.

\bibitem{edm57}
A.~R. Edmonds, Angular momentum in quantum mechanics, Princeton University
  Press, Princeton, 1957.

\bibitem{bau99}
A.~R. Bautista, G.~Co', A.~M. Lallena, Spin-orbit interaction in
  {H}artee-{F}ock calculations, Nuovo \ Cimento \ A 112 (1999) 1117.

\bibitem{co98b}
G.~Co', A.~M. Lallena, Tensor interaction in {H}artree-{F}ock calculations,
  Nuovo \ Cimento \ A 111 (1998) 527.

\bibitem{rin80}
P.~Ring, P.~Schuck, The nuclear many-body problem, Springer, Berlin, 1980.

\bibitem{co23}
G.~Co', Introducing the random phase approximation theory, Universe 9 (2023)
  141.

\bibitem{don14a}
V.~{De Donno}, G.~Co', M.~Anguiano, A.~M. Lallena, Coulomb and spin-orbit
  interactions in random-phase approximation calculations, Phys. \ Rev. \ C 89
  (2014) 014309.

\bibitem{don11b}
V.~{De Donno}, M.~Anguiano, G.~Co', A.~M. Lallena, Self-consistent continuum
  random phase approximation calculations of $^4${H}e electromagnetic
  responses, Phys. \ Rev. \ C 84 (2011) 037306.

\bibitem{don16}
V.~{De Donno}, G.~Co', M.~Anguiano, A.~M. Lallena, Self-consistent continuum
  random-phase approximation with finite-range interactions for charge-exchange
  excitations, Phys. \ Rev. \ C 93 (2016) 034320.

\bibitem{co16}
G.~Co', V.~{De Donno}, M.~Anguiano, A.~M. Lallena, Continuum random phase
  approximation with finite-range interactions, Eur. \ Phys. \ Jour. \ A 52
  (2016) 145.

\bibitem{dec80}
J.~Decharg\'e, D.~Gogny, {H}artree-{F}ock-{B}ogolyubov calculations with the
  {D1} effective interaction on spherical nuclei, Phys. \ Rev. \ C 21 (1980)
  1568.

\bibitem{ber91}
J.~F. Berger, M.~Girod, D.~Gogny, Time-dependent quantum collective dynamics
  applied to nuclear fission, Comp. \ Phys. \ Commun. 63 (1991) 365.

\bibitem{cea}
S.~Hilaire, M.~Girod, {H}artree-{F}ock-{B}ogoliubov results based on the
  {G}ogny force. {AMEDEE} database.
  \url{http://www-phynu.cea.fr/HFB-Gogny\_eng.htm}.

\bibitem{hil07}
S.~Hilaire, M.~Girod, Large-scale mean-field calculations from proton to
  neutron drip lines using the {D1S} {G}ogny force, Eur. \ Phys. \ J. \ A 33
  (2007) 237.

\bibitem{gor09}
S.~Goriely, S.~Hilaire, M.~Girod, S.~P\'eru, First
  {G}ogny-{H}artree-{F}ock-{B}ogoliubov nuclear mass model, Phys. \ Rev. \
  Lett. 102 (2009) 242501.

\bibitem{ang11}
M.~Anguiano, G.~Co', V.~{De Donno}, A.~M. Lallena, Phys. \ Rev. \ C 83 (2011)
  064306.

\bibitem{co85}
G.~Co', S.~Krewald, A model for particle emission induced by electron
  scattering, Nucl. \ Phys. \ A 433 (1985) 392.

\bibitem{led78}
C.~M. Lederer, V.~S. Shirley, Table of isotopes, 7th ed., John Wiley and sons,
  New York, 1978.

\bibitem{dej87}
C.~W. {De Jager}, C.~{De Vries}, Nuclear charge density distribution parameters
  from elastic electron scattering, At. \ Data \ Nucl. \ Data \ Tables 36
  (1987) 495.

\bibitem{bnlw}
{Brookhaven National Laboratory}, National nuclear data center.
  \url{http://www.nndc.bnl.gov/}.

\bibitem{ang13}
I.~Angeli, K.~P. Marinova, Table of experimental nuclear ground state charge
  radii: An update, Atomic \ Data \ and \ Nuclear \ Data \ Tables 99 (2013) 69.

\bibitem{hyd87}
C.~E. Hyde-Wright, et~al., Phys. \ Rev. \ C 35 (1987) 880.

\bibitem{kue83}
G.~K{\"u}chler, A.~Richter, E.~Spamer, W.~Steffen, W.~Kn{\"u}pfer, Nucl. \
  Phys. A 406 (1983) 473.

\bibitem{ahr75}
J.~Ahrens, et~al., Nucl. \ Phys. \ A 251 (1975) 479.

\bibitem{Langanke:1995he}
K.~Langanke, P.~Vogel, E.~Kolbe, {Signal for supernova muon-neutrino and
  tau-neutrino neutrinos in water Cherenkov detectors}, Phys. Rev. Lett. 76
  (1996) 2629--2632.
\newblock \href {http://arxiv.org/abs/nucl-th/9511032}
  {\path{arXiv:nucl-th/9511032}}, \href
  {https://doi.org/10.1103/PhysRevLett.76.2629}
  {\path{doi:10.1103/PhysRevLett.76.2629}}.

\bibitem{Rauscher:SPECTRUM}
T.~Rauscher, computer code spectrum, unpublished (2018).

\bibitem{smargd}
T.~Rauscher, computer code smaragd, version 0.9.3s, unpublished (2010 - 2022).

\bibitem{Rauscher:intjmodphyse}
T.~Rauscher, {The Path to Improved Reaction Rates for Astrophysics}, Int.\ J.
  Mod.\ Phys.\ E 20 (2011) 1071.

\bibitem{Jeukenne:1974zz}
J.~P. Jeukenne, A.~Lejeune, C.~Mahaux, {Optical-model potential in nuclear
  matter from Reid's hard core interaction}, Phys. Rev. C 10 (1974) 1391--1401.
\newblock \href {https://doi.org/10.1103/PhysRevC.10.1391}
  {\path{doi:10.1103/PhysRevC.10.1391}}.

\bibitem{Lejeune:1980zz}
A.~Lejeune, {Low-energy optical model potential in finite nuclei from Reid's
  hard core interaction}, Phys. Rev. C 21 (1980) 1107--1108.
\newblock \href {https://doi.org/10.1103/PhysRevC.21.1107}
  {\path{doi:10.1103/PhysRevC.21.1107}}.

\bibitem{McFadden:1966zz}
L.~McFadden, G.~R. Satchler, {Optical-model analysis of the scattering of 24.7
  MeV alpha particles}, Nucl. Phys. 84 (1966) 177--200.
\newblock \href {https://doi.org/10.1016/0029-5582(66)90441-X}
  {\path{doi:10.1016/0029-5582(66)90441-X}}.

\bibitem{Rauscher:2000fx}
T.~Rauscher, F.-K. Thielemann, {Astrophysical reaction rates from statistical
  model calculations}, Atom. Data Nucl. Data Tabl. 75 (2000) 1--351.
\newblock \href {http://arxiv.org/abs/astro-ph/0004059}
  {\path{arXiv:astro-ph/0004059}}, \href
  {https://doi.org/10.1006/adnd.2000.0834} {\path{doi:10.1006/adnd.2000.0834}}.

\bibitem{Fischer:2016cyd}
T.~Fischer, S.~Chakraborty, M.~Giannotti, A.~Mirizzi, A.~Payez, A.~Ringwald,
  {Probing axions with the neutrino signal from the next galactic supernova},
  Phys. Rev. D 94~(8) (2016) 085012.
\newblock \href {http://arxiv.org/abs/1605.08780} {\path{arXiv:1605.08780}},
  \href {https://doi.org/10.1103/PhysRevD.94.085012}
  {\path{doi:10.1103/PhysRevD.94.085012}}.

\bibitem{Lunardini:2004bj}
C.~Lunardini, A.~Y. Smirnov, {Neutrinos from SN1987A: Flavor conversion and
  interpretation of results}, Astropart. Phys. 21 (2004) 703--720.
\newblock \href {http://arxiv.org/abs/hep-ph/0402128}
  {\path{arXiv:hep-ph/0402128}}, \href
  {https://doi.org/10.1016/j.astropartphys.2004.05.005}
  {\path{doi:10.1016/j.astropartphys.2004.05.005}}.

\bibitem{Fogli:2004ff}
G.~L. Fogli, E.~Lisi, A.~Mirizzi, D.~Montanino, {Probing supernova shock waves
  and neutrino flavor transitions in next-generation water-Cerenkov detectors},
  JCAP 04 (2005) 002.
\newblock \href {http://arxiv.org/abs/hep-ph/0412046}
  {\path{arXiv:hep-ph/0412046}}, \href
  {https://doi.org/10.1088/1475-7516/2005/04/002}
  {\path{doi:10.1088/1475-7516/2005/04/002}}.

\bibitem{Fiorillo:2023frv}
D.~F.~G. Fiorillo, M.~Heinlein, H.-T. Janka, G.~Raffelt, E.~Vitagliano,
  R.~Bollig, {Supernova simulations confront SN 1987A neutrinos}, Phys. Rev. D
  108~(8) (2023) 083040.
\newblock \href {http://arxiv.org/abs/2308.01403} {\path{arXiv:2308.01403}},
  \href {https://doi.org/10.1103/PhysRevD.108.083040}
  {\path{doi:10.1103/PhysRevD.108.083040}}.

\bibitem{Lucente:2022vuo}
G.~Lucente, L.~Mastrototaro, P.~Carenza, L.~Di~Luzio, M.~Giannotti, A.~Mirizzi,
  {Axion signatures from supernova explosions through the nucleon
  electric-dipole portal}, Phys. Rev. D 105~(12) (2022) 123020.
\newblock \href {http://arxiv.org/abs/2203.15812} {\path{arXiv:2203.15812}},
  \href {https://doi.org/10.1103/PhysRevD.105.123020}
  {\path{doi:10.1103/PhysRevD.105.123020}}.

\bibitem{ost92}
F.~Osterfeld, Nuclear spin and isospin excitations, Rev. \ Mod. \ Phys. 64
  (1992) 491.

\bibitem{ang06}
M.~Anguiano, G.~Co', A.~M. Lallena, Proton emission induced by polarized
  photons, Phys. \ Rev. \ C 74 (2006) 044603.

\end{thebibliography}

\newpage
%
% figure density
%-----------------------------------------

% spectrum
%-----------------------------------------
% M2

% e e' M4
%-----------------------------------------

%-----------------------------------------

% photoabsorption
%-----------------------------------------

% D1M axion log
%-----------------------------------------

%-----------------------------------------
% axion linear
%-----------------------------------------

%%%%%%%%%%%%%%%%%%%%%%%%%%%%%%%%%%
%%%%%%%%%%%%%%%%%%%%%%%%%%%%%%%%%%%%%%%%

%%%%%%%%%%%%%%%%%%%%%%%%%%%%%%%%%%
%%%%%%%%%%%%%%%%%%%%%%%%%%%%%%%%%%%%%%%%

\end{document}